\pdfoutput=1
\documentclass[fleqn,usenatbib]{mnras}

\usepackage{newtxtext,newtxmath}

\usepackage[T1]{fontenc}

\DeclareRobustCommand{\VAN}[3]{#2}
\let\VANthebibliography\thebibliography
\def\thebibliography{\DeclareRobustCommand{\VAN}[3]{##3}\VANthebibliography}

\usepackage{graphicx}	
\usepackage{amsmath}	
\usepackage{float}
\usepackage{footnote}
\usepackage[para,online,flushleft]{threeparttable}

\title[Flares Study of PKS B1222+216 in 2014]{Temporal and Spectral Study of PKS B1222+216 Flares in 2014}

\author[A. Chatterjee et al. 2021]{
Anshu Chatterjee,$^{1}$\thanks{E-mail: anshuchatterjee90@gmail.com}
Abhradeep Roy,$^{1}$
Arkadipta Sarkar$^{1}$
and Varsha R. Chitnis$^{1}$
\\
$^{1}$Department of High Energy Physics, Tata Institute of Fundamental Research, Colaba, Mumbai-400005, India\\
}

\date{Accepted 17 September 2021. Received 16 September 2021; in original form 23 May 2021}

\pubyear{2021}
\makesavenoteenv{table}
\makesavenoteenv{tabular}

\begin{document}
\label{firstpage}
\pagerange{\pageref{firstpage}--\pageref{lastpage}}
\maketitle

\begin{abstract}
We report on a temporal and spectral study of a flat spectrum radio quasar, PKS B1222+216, in a flare state to get insight into the acceleration and emission mechanisms inside the jet. It is one of the brightest and highly active blazars in the MeV-GeV regime. The long term multi-waveband light curves of this object showed a flaring activity in 2014 with two distinct flares. The work presented here includes the study of flux-index variation, flare fitting, hardness ratio, and the spectral modelling of X-ray and $\gamma-$ray data. The flux-index correlation found in the MeV-GeV regime indicates a ‘softer when brighter’ feature. The modelling of $\gamma-$ray light curves suggests that the low energy particles initiate both the flares followed by the injection of the high energy particles. The short rise time indicates the presence of Fermi first-order acceleration. A single-zone leptonic model is used to fit the multi-waveband spectral energy distributions generated for both flares. The spectral energy distribution modelling shows that the inverse Compton scattering of the photon field reprocessed from the Broad Line Region primarily accounts for the GeV emission. In addition, we have reported a shift in the break-energy in the soft X-ray regime during flares, which is due to a rapid change in the injection spectrum.
\end{abstract}

\begin{keywords}
Galaxies: active -- Galaxies: jets -- Radiation mechanisms: non-thermal -- Galaxies: individual: PKS B1222+216
\end{keywords}



\section{Introduction}
Blazars are generally considered one of the classes of the most violently variable objects over the entire electromagnetic range in the universe. Blazars form a special kind of radio-loud active galactic nuclei (AGN) with jet emission oriented at small angles ($\leq 10^\circ$) to the observer's line of sight \citep{Urry1995PASP}. The relativistic beaming effect of jet emission along the line of sight provides a self-explanation of its violent nature \citep{Blandford1979ApJ}. Variability over a time scale of minutes to weeks is a common phenomenon for these objects and is observed across the entire electromagnetic spectrum. Based on the presence of strong emission lines in the optical spectrum, blazars are classified into two main categories: BL Lacertae (BL Lac) objects and flat-spectrum radio quasars (FSRQs). BL Lacs are objects with weak emission lines in their optical spectra, whereas the presence of emission lines of equivalent widths (EWs) greater than 5 {$\mathring{\text{A}}$} is general criteria for FSRQ \citep{Urry1995PASP, Sbarrato2012MNRAS}. Presently, a total of $22$ BL Lacs and $43$ FSRQs have been identified with energies above $100$ GeV by the Fermi Large Area Telescope \citep[LAT,][]{Abdollahi2020ApJS}.

The spectral energy distribution (SED) of an FSRQ shows a characteristic double hump structure. Leptonic or hadronic models can explain the origin of this double hump structure. According to both models, the first hump comes from the synchrotron emission of electrons accelerated to ultra-relativistic energies inside the emission region. The explanation of the second hump is different in different models. In the case of a hadronic model, the relativistic protons having energies above the interaction threshold produce a high energy hump either via pion photoproduction mechanism or via proton synchrotron radiation
\citep{2001APh....15..121M,2003APh....18..593M}. Whereas, in a leptonic model, it is generally assumed that the high-energy hump is a result of the up-scattering of the low energy seed photons via inverse Compton (IC) scattering by the primary electrons accelerated in the relativistic jet. The source of the seed photon field can be internal and/ or external to the emission zone. If the synchrotron photons, generated from the same electron population that takes part in IC scattering, then resultant emission is called synchrotron self-Compton \citep[SSC,][]{Bloom1996ApJ}. On the other hand, if an external photon field is involved in this process, then the corresponding emission is called external Compton (EC). The possible sources of the external photons are direct thermal photons from the accretion disk \citep[EC-disk,][]{Dermer1992A&A, Dermer1993ApJ}, or reprocessed photons either from the broad-line region (BLR) or from the dusty torus \citep[DT i.e. EC-BLR and EC-DT,][]{Ghisellini1998MNRAS}. In this work, we have considered a leptonic scenario to explain the observed SED.
PKS B1222+216 \citep[4C +21.35; z = 0.432,][]{Osterbrock1987ApJ,Abdo2010ApJ} is one of the brightest FSRQ detected in very high energy (VHE) $\gamma-$ray regime. PKS B1222+216 was first detected by the very long baseline interferometry (VLBI) observation which showed an asymmetric nature in radio structure on milliarcsec scales \citep{Hooimeyer2010MNRAS}. Further radio observations showed that the flux ratio of the core portion to the extended structure in radio wavelength is of the order of unity and slightly less luminous compared to its large-scale structure, which is rare in blazars. Based on these facts, PKS B1222+216 was formally categorized as \lq lobe dominated\rq \citep{Sbarrato2012MNRAS}. The source was detected in $\gamma$-rays for the first time by Energetic Gamma Ray Experiment Telescope (EGRET) onboard the Compton Gamma Ray Observatory \citep{Hartman1999ApJS}. PKS B1222+216 was present in the first LAT catalog \citep{Abdo2010ApJS} with the name 1FGL J1224.7+2121 and since then it has been very active in the GeV regime with some occasional brightness enhancements \citep{Tanaka2011ApJ}. The first outburst was reported by the Fermi-LAT collaboration in April 2009 \citep{Longo2009ATel}. In 2010, PKSB1222+216 went through some major flaring activities ($\sim$ $10^{-5}\text{ph}\text{ cm}^{-2}\text{ s}^{-1}$). The two distinct flares were observed in April and June 2010 \citep{Tanaka2011ApJ}. The second flare in June was much more violent and associated with the VHE observation from MAGIC with the shortest flux doubling time scale of $\sim10$ min \citep{Mariotti2010Atel,2011ApJ...730L...8A}. This detection made PKS B1222+216 the third FSRQ detected at VHE regime after 3C 279 and PKS 1510-089 \citep{MAGIC2008AIPC,HESS2010HEAD,2011ApJ...730L...8A}. For this giant flare, there were no simultaneous observations in the X-ray/UV band. But there were some quasi-simultaneous observations in the decaying phase in optical \citep{Dominici2010ATel} and UV/X-ray bands \citep{Verrecchia2010ATel}. In 2014, the object showed some flaring activities in the GeV regime but the flux strength was an order of magnitude less compared to 2010. This was earlier reported in \citet{Verrecchia2014ATel}. A closer examination revealed that the flaring activity in 2014 was a combination of two flares which were detected in both X-ray and GeV bands. In this work, we mainly focus on both the flares of PKS B1222+216 in 2014. In \citet{Bhattacharya2021MNRAS}, this activity is mentioned as a moderate-activity state (MS) and only a small portion ($\sim7$ days) of the second flare was modelled with one zone leptonic scenario. In this work, we have studied both flares separately and presented spectral characterization and broadband modelling of 2014 flare in great detail. For both the flares, we have analyzed the $\gamma-$ray data in three consecutive bands with the shortest possible timescale and studied correlations between them. We report a detailed study of X-ray data which reveals the shifting of break-energy during the course of the flaring activity. For further study, one flare is divided into multiple blocks and broadband SED modelling with leptonic scenario has been carried out for each block to understand the underlying emission mechanisms. Due to lack of flux variability, another flare has been modelled as a single SED. Details of the analysis of multi-wavelength data used in this work are described in \autoref{sec:obs}. In \autoref{sec:ana}, we present our temporal and spectral analysis as well as the broadband SED modelling, followed by a discussion and conclusions in \autoref{sec:dis} and \autoref{sec:con} respectively.

\section{Observations and Data Reduction}
\label{sec:obs}
For this study, we used Fermi-LAT data, spanning 118 days, for both spectral analysis and broadband modelling of two flare states. For broadband modelling, X-ray and UV data are taken from instruments onboard the Neil Gehrels Swift Observatory. We have also used publicly available spectroscopic data from the SPOL-CCD of the Steward Observatory in this work. In this section, we describe the data extraction and analysis procedures used.

\subsection{SPOL-CCD of Steward
Observatory: Optical Data}
SPOL-CCD is a Spectropolarimeter instrument at Steward Observatory at the University of Arizona. It is a combination of polarimeter and transmission-optics spectrograph into a self-contained, portable instrument. SPOL provides data both in R ($6520$ {\AA}) and V ($5517$ {\AA}) band in terms of magnitudes.
The details of the instrument and data analysis procedures are given in \citep{Smith2009}. As a part of the Fermi multi-wavelength support program, SPOL monitored bright Fermi sources regularly from 2008 to 2019. 
Publicly available V and R band spectroscopic data are obtained from SPOL data base\footnote{\url{http://james.as.arizona.edu/~psmith/Fermi/}}. Obtained magnitudes were corrected for Galactic reddening and extinction using online resources\footnote{\url{https://irsa.ipac.caltech.edu/applications/DUST/}} supported by NASA.

\subsection{\textit{Swift}-UVOT: Optical-UV Data}
\textit{Swift}-UVOT or Ultraviolet/Optical Telescope \citep{Roming2005ssr} is one of the three instruments onboard space-based Neil Gehrels \textit{Swift} Observatory. It consists of three optical filters i.e. V, B, U and three UV filters i.e. UVW1, UVM2, and UVW2. The data corresponding to each filter are publicly available from \textit{Swift} database\footnote{\href{https://heasarc.gsfc.nasa.gov/docs/archive.html}{https://heasarc.gsfc.nasa.gov/docs/archive.html}}. All data available within our selected periods were integrated using dedicated software tool UVOTIMSUM distributed within HEASOFT package (v6.27.2)\footnote{\href{https://heasarc.gsfc.nasa.gov/docs/software/heasoft/}{https://heasarc.gsfc.nasa.gov/docs/software/heasoft/}}. For extracting photon counts from integrated images, source regions of radii $5$" and $10$" were chosen around the source location for optical and UV filters respectively. For background data, a circular region of radius $25$" was chosen excluding source position. The tool UVOTSOURCE was used to obtain the final magnitudes for each filter. The observed magnitudes were corrected for galactic extinction and converted into flux using zero-point magnitudes \citep{Poole2008MNRAS}.

\subsection{\textit{Swift}-XRT: Soft X-ray Data}
\textit{Swift}-XRT or X-ray Telescope \citep{Burrows2005ssr} is another instrument onboard the Neil Gehrels \textit{Swift} Observatory which monitors sky in soft X-ray regime in the energy range of $0.3$–$8.0$ keV. Publicly available soft X-ray data were obtained from the \textit{Swift} archive\footnote{\href{https://www.swift.ac.uk/}{https://www.swift.ac.uk/}}. The data were analyzed using dedicated XRT data analysis software (XRTDAS) distributed within the HEASOFT package. The cleaning of raw event files
has been done using the XRTPIPELINE-0.13.5 tool, and the
XRTPRODUCTS-0.4.2 tool was used to generate the spectral files. The spectra of individual observations were combined using the ADDSPEC tool and rebinned with a minimum of 20 photons per bin with GRPPHA (v3.1.0). The final XRT spectra were then fitted with different spectral models (power law and broken power-law) using XSPEC (v12.11.0) tool. These modellings include the interstellar absorption of low-frequency end in terms of neutral hydrogen column density \citep{Kalberla2005aap}. The power-law model is given as,
\begin{equation}
    \frac{dN}{dE} = K E^{-\Gamma_1}
\end{equation}
where $\Gamma_1$ is the spectral index. The broken power law is defined as,
\begin{equation}
    \frac{dN}{dE} =
    \begin{cases}
    K E^{-P_1} & \text{ when } E \leq E_{b} \\
    K E^{P_2-P_1}_{b} (E/1\text{ keV})^{-P_2} & \text{ when } E > E_b
    \end{cases}
\end{equation}
where $P_1$ and $P_2$ represent the spectral indices before and after the break-energy $E_b$ respectively. $K$ is the normalization constant.

\subsection{Fermi LAT: GeV Data}
Fermi-LAT \citep{Atwood2009ApJ} is one of the two $\gamma$-ray detectors onboard the Fermi Gamma-ray Space Telescope (Fermi) mission. Fermi-LAT is a pair-conversion $\gamma$-ray telescope, with a field of view (FoV) of above 2 sr, operating in the energy range from $20$ MeV to $300$ GeV. It is the most sensitive instrument available in this energy range \citep{Ackermann2012ApJs}. Fermi-LAT operates in all-sky survey mode and scans the whole sky in $3$ hours \citep{Atwood2009ApJ}. In this work, we used PASS8 Fermi-LAT data of PKS B1222+216 covering MJD $56680-56748$ (2014-01-23 to 2014-04-01) and MJD $56950-57000$ (2014-10-20 to 2014-12-09). To avoid the comparatively large point spread function at low energy regime, we analyzed data over the energy range of $0.1-300$ GeV\footnote{\href{https://fermi.gsfc.nasa.gov/cgi-bin/ssc/LAT/LATDataQuery.cgi}{https://fermi.gsfc.nasa.gov/cgi-bin/ssc/LAT/LATDataQuery.cgi}}. For this analysis, we used the FERMITOOLS\footnote{\href{https://fermi.gsfc.nasa.gov/ssc/data/analysis/software/}{https://fermi.gsfc.nasa.gov/ssc/data/analysis/software/}} software package version v11r5p3 and user-contributed ENRICO software \citep{Sanchez2013arxiv}. The GTSELECT tool was used for data selection and quality checks. To select good time intervals, GTMKTIME tool used a filter \lq\lq \texttt{DATA\_QUAL>0}\rq\rq $\&\&$ \lq\lq \texttt{LAT\_CONFIG==1}\rq\rq. To exclude the background $\gamma$-ray events from the earth's limb i.e events from the interactions of cosmic rays with the ambient matter, a maximum zenith angle cut of $90^\circ$ was set. To generate flux and SED data points, we used the maximum likelihood optimization \citep{Abdo2009ApJS}. The analysis included source events from a circular region of $15^{\circ}$ radius around PKS B1222+216, which is called the region of interest (ROI). The fourth Fermi-LAT catalog (4FGL catalog: \citep{Thompson2019HEAD, Abdollahi2020ApJS} was used to include the contributions from all available sources inside the ROI. The instrument response function \texttt{P8R3\_SOURCE\_V2} was used in this analysis. The background model consists of two components, one accounts for galactic diffused emission, \texttt{gll\_iem\_v07.fits} and other accounts for isotropic background emission, \texttt{iso\_P8R3\_SOURCE\_V2\_v1.txt}. The spectral model of the source was considered as log-parabola as mentioned in the 4FGL catalog. In the likelihood optimization process, spectral
parameters of all the sources including PKS B1222+216 inside the ROI were kept
free, whereas the spectral parameters of the sources beyond $15^{\circ}$ from PKS B1222+216 were kept fixed to the values according to the 4FGL catalog. The unbinned likelihood\footnote{\href{https://fermi.gsfc.nasa.gov/ssc/data/analysis/scitools/likelihood_tutorial.html}{https://fermi.gsfc.nasa.gov/ssc/data/analysis/scitools}} method was used to obtain the detection significance of the sources. The Test Statistic is defined as $\text{TS} = -2[ln(L_0)-ln(L_1)]$, where $L_0$ is the maximum likelihood value for a model without an additional source (i.e. 'null hypothesis') and $L_1$ is the maximum likelihood value for a model with the additional source at a specified location. The repetitive likelihood analysis excludes all sources with TS values less than $9$ which corresponds to a detection significance of $3\sigma$ \citep{Mattox1996ApJ}. To obtain the SED points, we divided the whole energy range into a suitable number of energy bins (5-6)
and repeated the likelihood analysis with a fixed value of global fit parameters. Similarly, to produce Fermi-LAT light curve, we have divided the whole period into multiple bins and performed the likelihood analysis for each time bin. 
For energy bins with the low test statistics value i.e. less than 9, upper limits were calculated with $95\%$ confidence level, and an assumed systematic error of $30\%$ using the method described in \citep{Rolke2005}. The effect of absorption of $\gamma-$ray photons due to the extragalactic background light (EBL) was taken into account in the EBL model.

\section{Analysis and Results}
\label{sec:ana}
\begin{figure}
\centering
\includegraphics[width=0.48\textwidth]{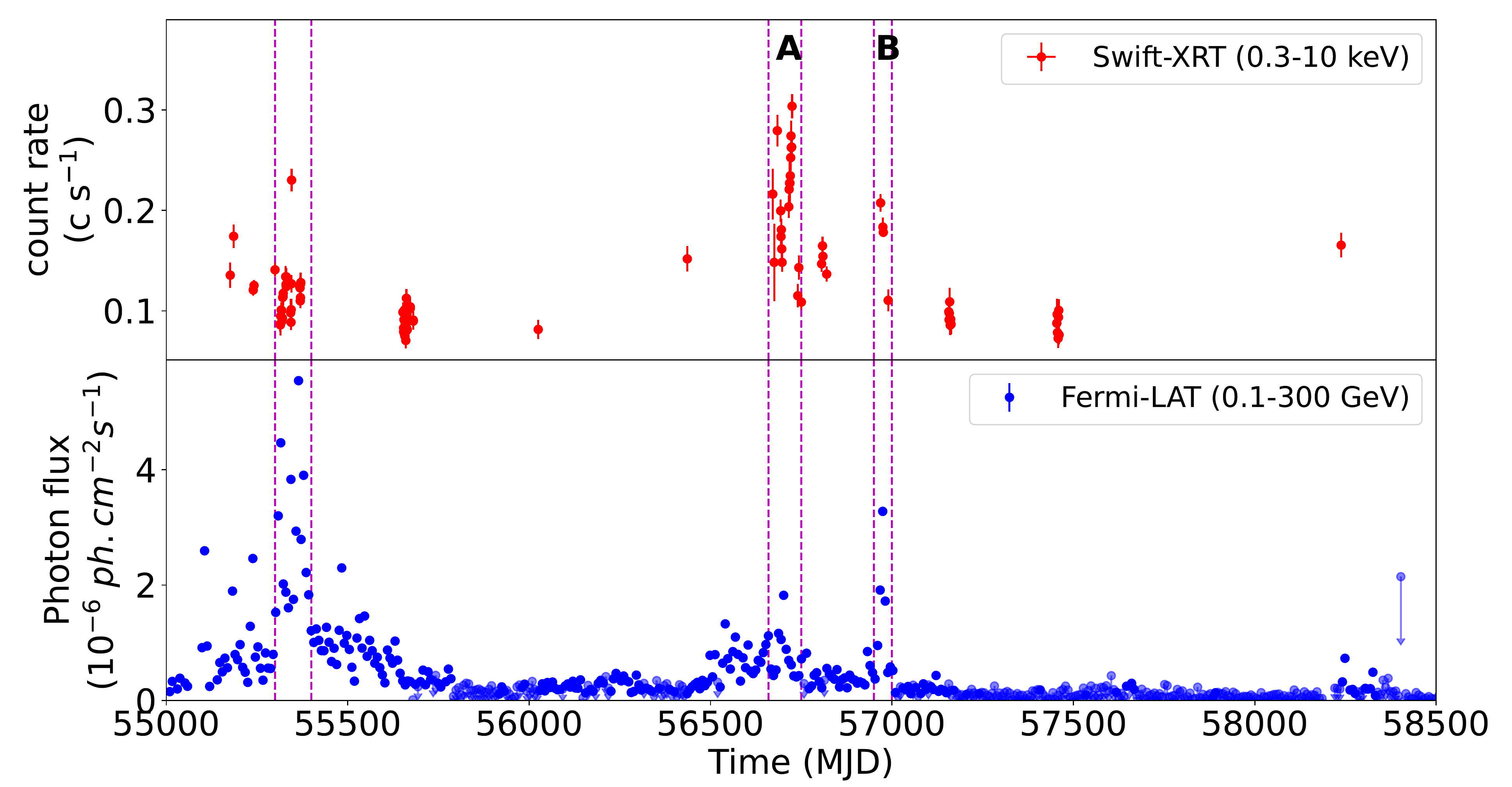}
\caption{Long term X-ray and $\gamma-$ray LCs of PKS B1222+216 over MJD $54800$ - $58400$. Upper panel represents \textit{Swift}-XRT LC over energy band $0.3 - 10$ keV generated by UKSSDC web-tool. The lower panel represents long term Fermi-LAT weekly binned LC based on data from aperture photometry. Based on available data, three possible flare periods are marked with red dashed lines.}
\label{fig:XRT-Fermi-longterm-LC}
\end{figure}

\begin{figure}
\centering
\begin{tabular}{c}
\includegraphics[trim={0 0 0 0},clip, width=0.47\textwidth]{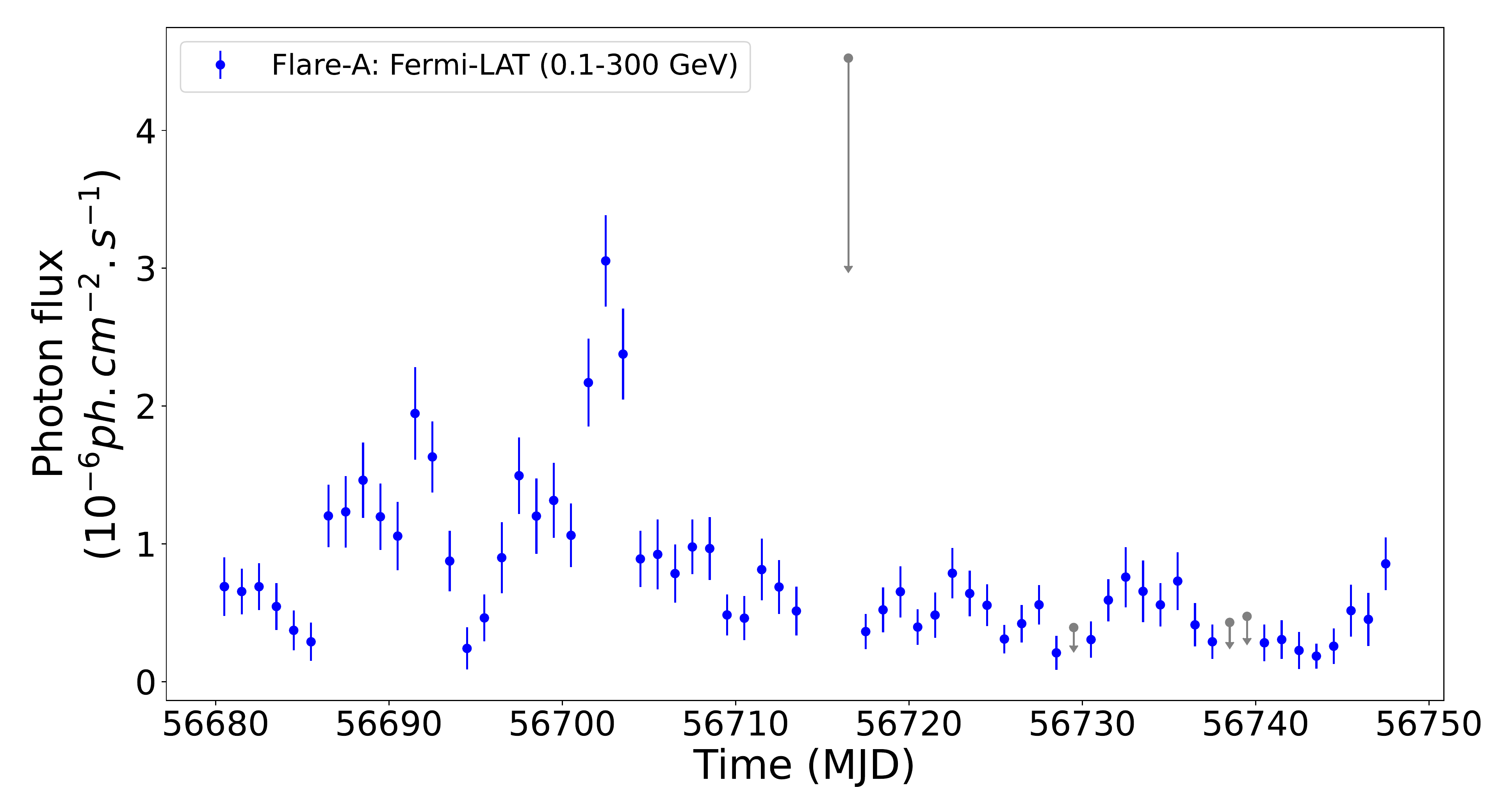}\\
\textbf{(a)}  \\[6pt]
\end{tabular}
\begin{tabular}{c}
\includegraphics[trim={0 0 0 0},clip,width=0.47\textwidth]{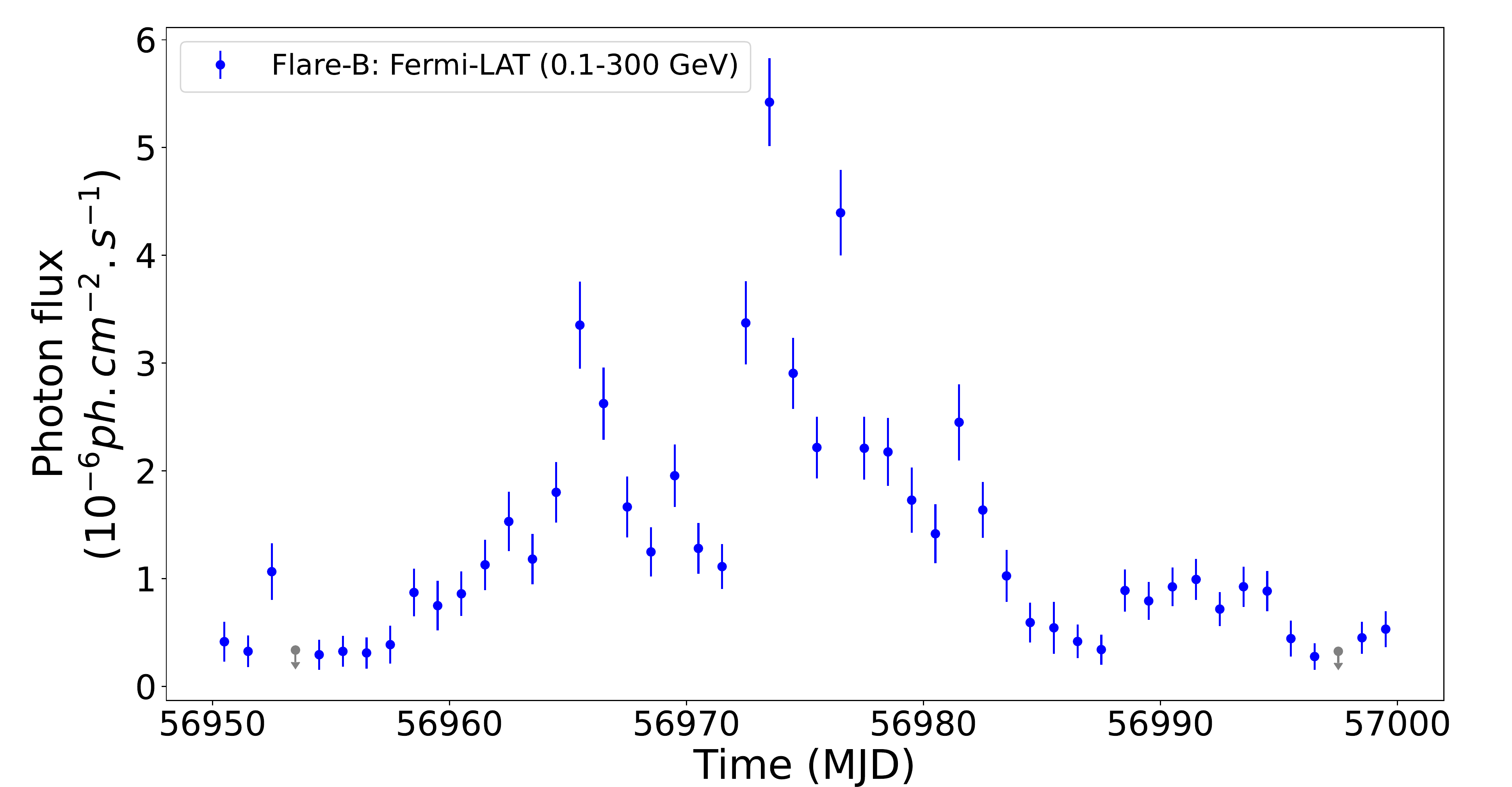}\\
\textbf{(b)}  \\[6pt]
\end{tabular}
\caption{Daily binned Fermi-LAT ($0.1$-$300$ GeV) light curve of PKS B1222+216 for both flare periods in $2014$ (see  \autoref{tab:flare_details}). Panel \textbf{(a)} and \textbf{(b)} represent the light curve for Flare-A and B respectively. The minimum TS criteria for data point is chosen as $9$ (i.e. $3\sigma$). The grey points with downward arrows represent the upper limit in the corresponding time bin.}
\label{fig:dailly-LC}
\end{figure}

Based on Fermi-LAT and \textit{Swift}-XRT long term light curves (\autoref{fig:XRT-Fermi-longterm-LC}), three flare periods of PKS B1222+216 were found, one in $2010$ (leftmost region marked by
dashed lines) and other two in the first and fourth quarter of $2014$
(middle and rightmost regions marked with dashed lines respectively). The \textit{Swift}-XRT and Fermi-LAT data were taken from UKSSDC web-tool\footnote{\href{https://www.swift.ac.uk/user_objects/}{https://www.swift.ac.uk/user\_objects/}} and aperture photometry database\footnote{\href{https://fermi.gsfc.nasa.gov/ssc/data/access/lat/msl_lc/}{https://fermi.gsfc.nasa.gov/ssc/data/access/lat/msl\_lc/}} respectively.
All of these periods were characterized by $\gamma-$ray  flux $F_{0.1-300 \text{ GeV}} \geq 5 \times 10^{-7} \text{ph cm}^{-2} \text{s}^{-1}$ and also represented rapid simultaneous flux variation in both the GeV and soft X-ray regimes.
The leftmost flare period of \autoref{fig:XRT-Fermi-longterm-LC} represents the giant flare states in 2010 which had already been studied in great detail in both GeV and TeV regimes \citep{Tavecchio2011A&A}. In the present work, we concentrate on two flares seen in 2014. The details of both flares are given in \autoref{tab:flare_details}.
\begin{table}
	\centering
	\caption{Details of two chosen flare states of PKS B1222+216}
	\label{tab:flare_details}
	\begin{tabular}{|l|c|c|c|} 
		\hline
		 & MJD & YYYY-MM-DD & Duration (days)\\
		\hline
		flare-A & 56680 - 56748 & 2014-01-23 to 2014-04-01 & 68\\
		\hline
		flare-B & 56950 - 57000 & 2014-10-20 to 2014-12-09 & 50\\
		\hline
	\end{tabular}
\end{table}
The Fermi-LAT light curve (LC) over full energy range ($0.1-300$ GeV) with binning of one day are extracted for both the flare periods and depicted in \autoref{fig:dailly-LC}. The data points are produced with the criteria of the minimum significance of $3\sigma$. For time bins where the minimum significance criteria are not satisfied, upper limits are provided with a confidence level of $95\%$. In the following LC analysis, we have considered only data points and omitted the upper limits from the corresponding time bins. The close inspection shows that both LCs contain multiple subflare structures within them. Based on the flux variability and the peak GeV flux, Flare-B is found to be more active and violent compared to Flare-A. On the other hand, flux variability in the soft X-ray regime is the most interesting feature of Flare-A.
\subsection{Flux-Index Variation}
\begin{figure*}
    \centering
	\includegraphics[width=0.49\linewidth]{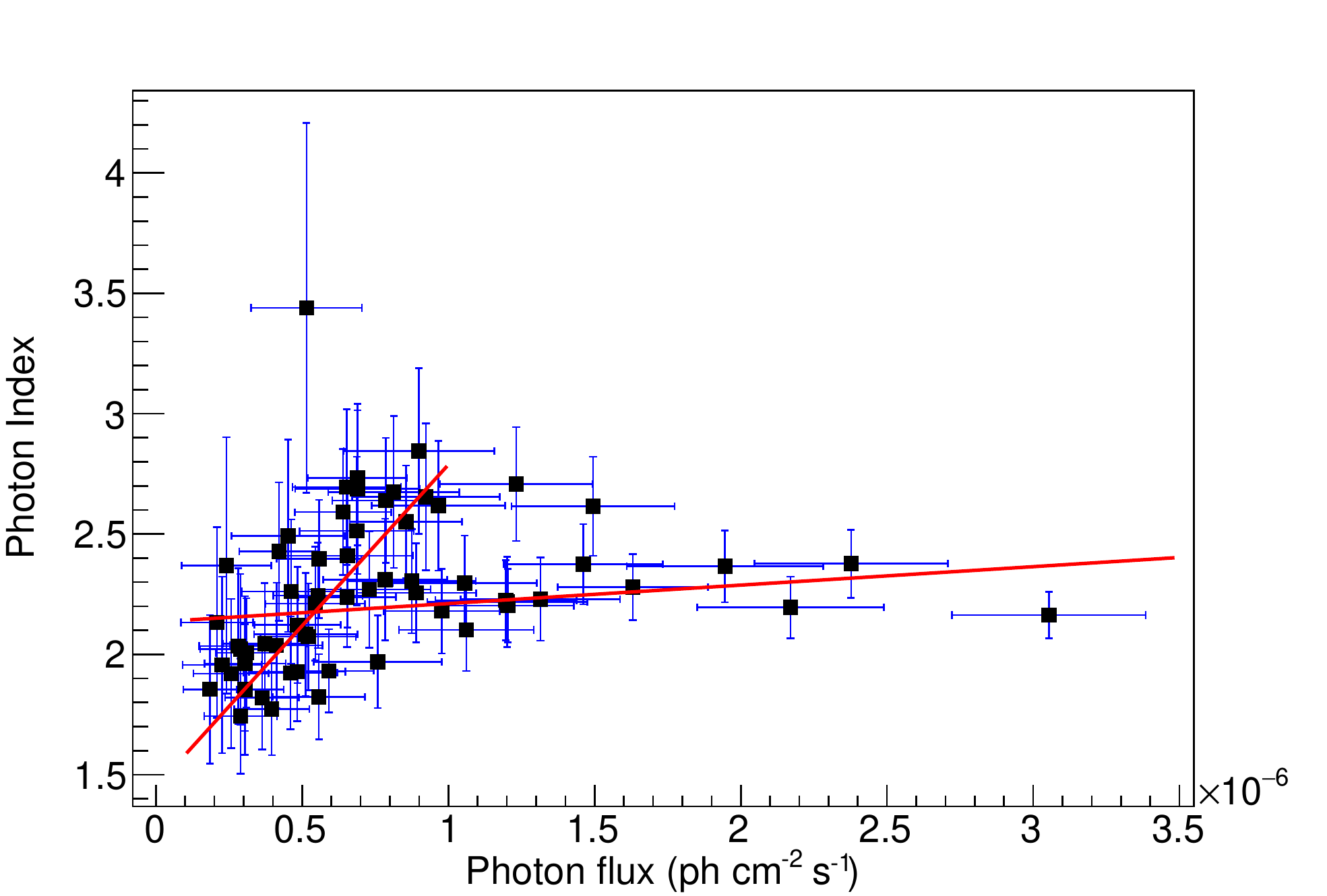}
	\includegraphics[width=0.49\linewidth]{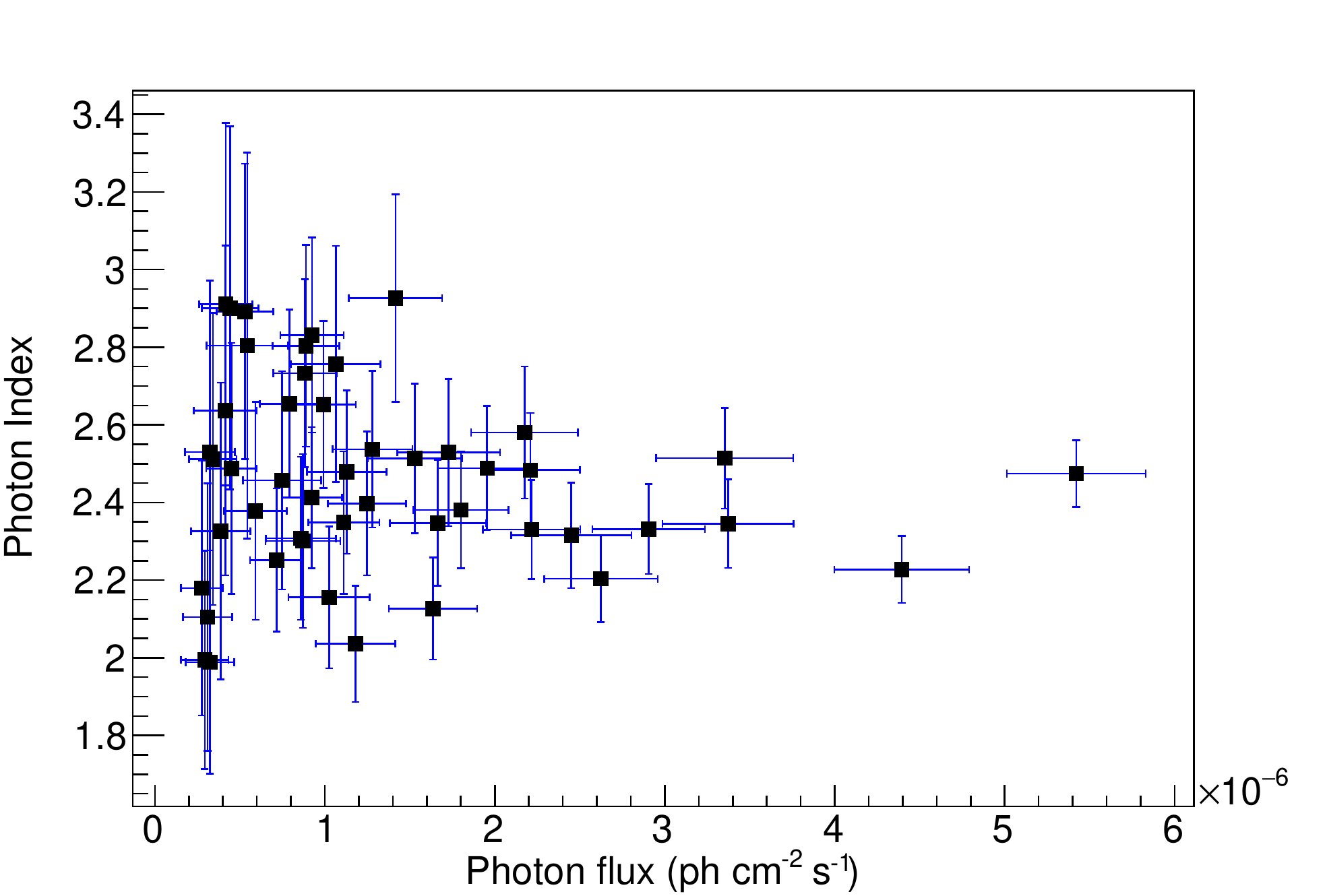}
	\caption{
	Flux-Index correlation showing a daily variation of photon index as a function of $\gamma-$ray flux for Flare-A (left panel) and Flare-B
    (right panel). In the left panel, data are fitted with a simple straight line of the form Mx+C. Fits carried out for the entire data as well as for
    data points with flux below $10^{-6} \text{ ph cm}^{-2} \text{s}^{-1}$ are shown.}
	\label{fig:flux-index corr}
\end{figure*}

\begin{figure*}
    \centering
	\includegraphics[width=0.49\linewidth, height=6.6cm]{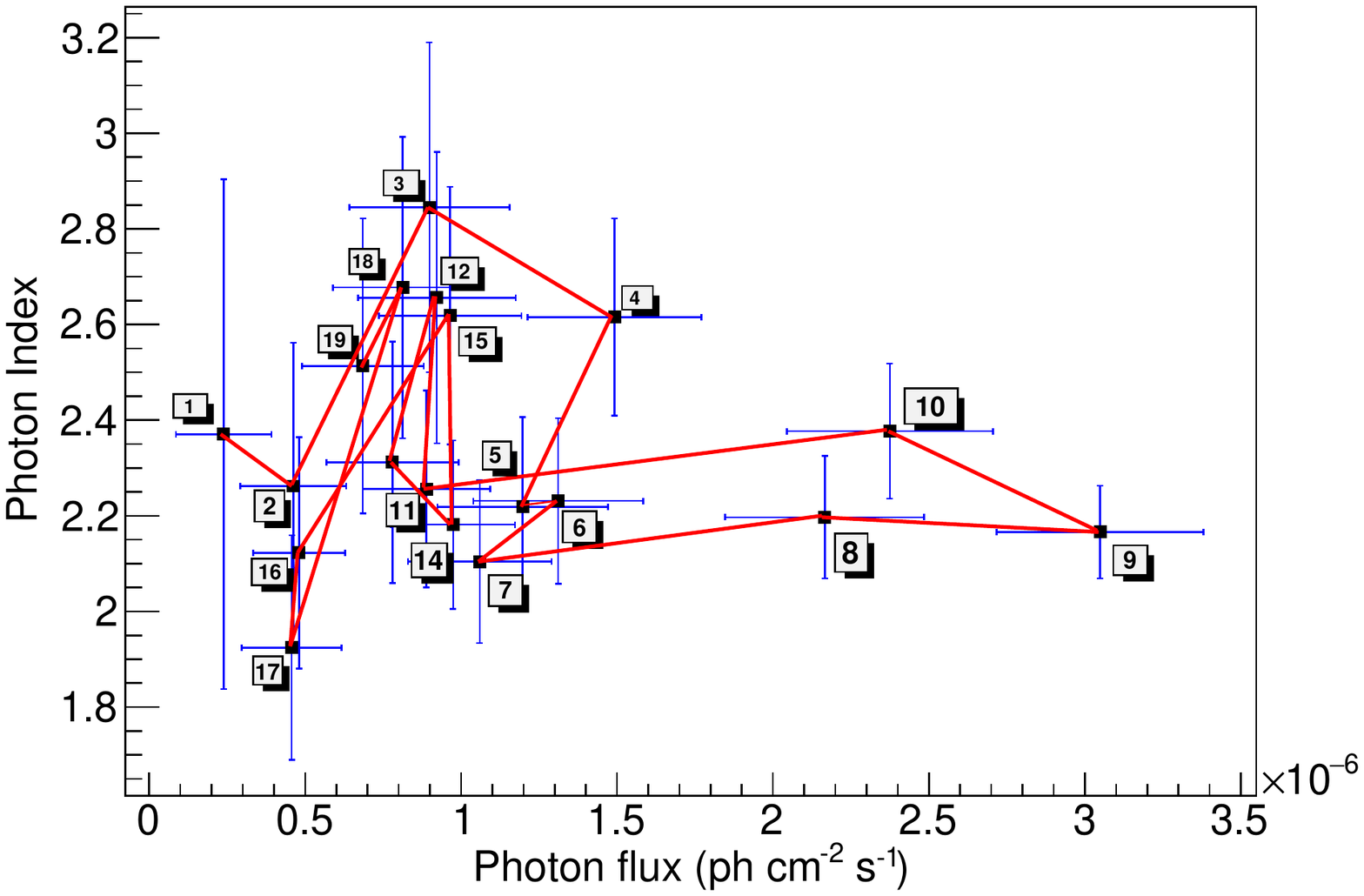}
	\includegraphics[width=0.49\linewidth]{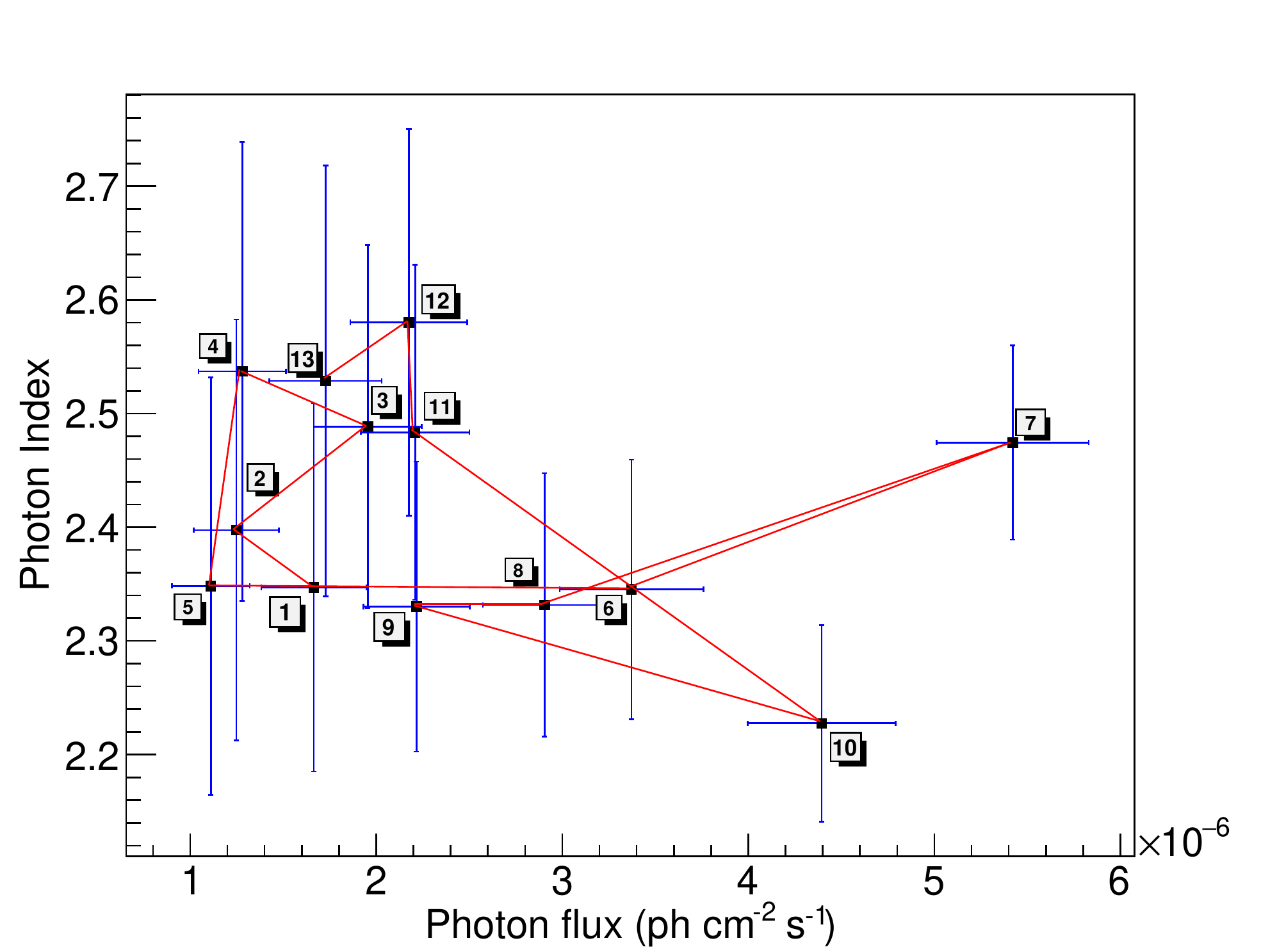}
	\caption{
    Flux-Index hysteresis:  Left and right panels represent the plots for Flare-A and Flare-B respectively.}
	\label{fig:flux-index hys}
\end{figure*}

To understand the change in the particle spectrum, we have extensively studied the flux-index correlation in the $\gamma$-ray band. In the process of Fermi LC extraction, we have kept the photon-index free, and thus the likelihood method determines its best-fitted value in each time bin. \autoref{fig:flux-index corr} represents flux-index correlation plots in the $\gamma-$ray band, which show the variation of photon index as a function of integral photon flux. The left and right panels of \autoref{fig:flux-index corr} represent the correlation plot for Flare-A and B, respectively. The scattered distribution of data points in the right figure represents very little flux-index correlation in Flare-B. On the other hand, the accumulation of data points in the \lq low flux low index\rq ~region provides the possibility of correlation in Flare-A, which we discuss later. For a further detailed study, we have considered a small portion of the flare around peak flux, where the flux is greater than $10\%$ of the peak flux. This criteria provides two short time windows i.e. MJD $56694$ -- $56713$ ($19$ days) and MJD $56967$ -- $56980$ ($13$ days) for Flare-A and B respectively. \autoref{fig:flux-index hys} shows the correlation plot over these short periods where points are marked with numbers indicating days from the beginning. The right panel shows the flux-index hysteresis plot for Flare-A and the left one for Flare-B. In the case of Flare-A, clear anticlockwise trends can be observed between $6$th to $11$th days from starting.
\begin{figure*}
\centering
\begin{tabular}{c}
\includegraphics[trim={1.0cm 0 1.35cm 0},clip, width=0.94\textwidth]{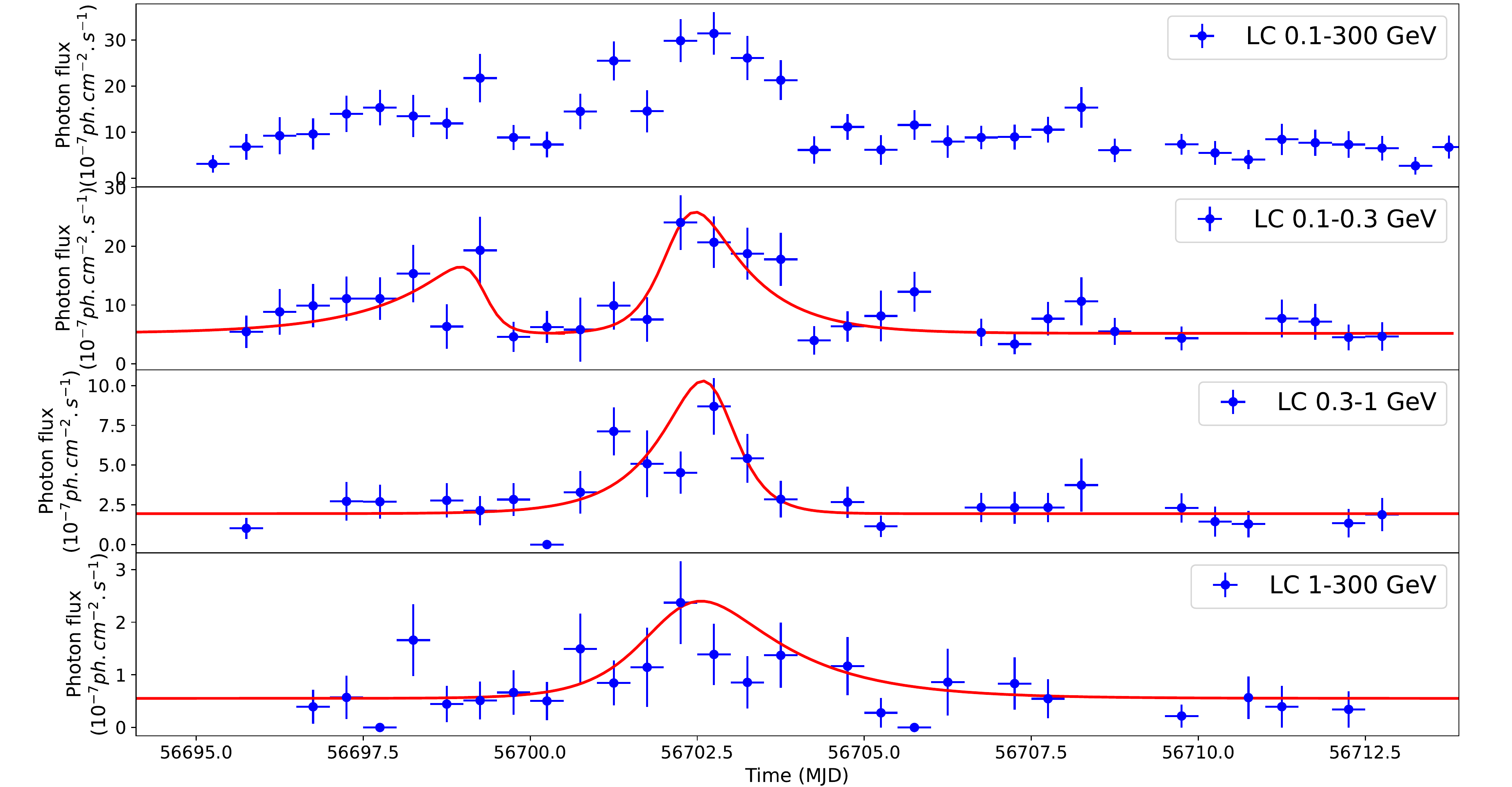}\\
\textbf{(a)}  \\[6pt]
\end{tabular}
\begin{tabular}{c}
\includegraphics[trim={1.0cm 0 1.35cm 0},clip, width=0.94\textwidth]{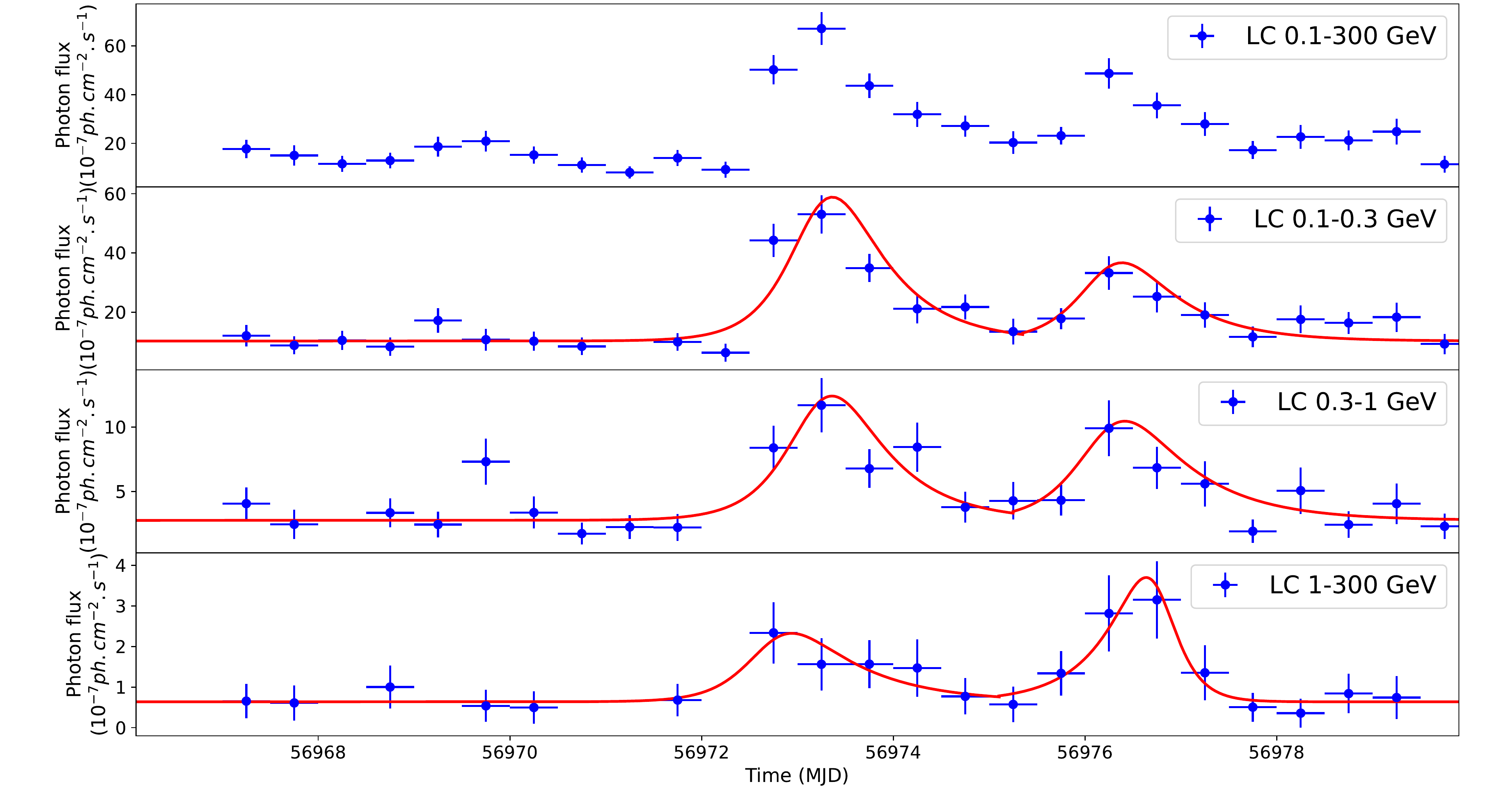} \\
\textbf{(b)}  \\[6pt]
\end{tabular}
\caption{12 hr binned $\gamma$-ray LC of PKS B1222+216 during selected period of Flare-A (MJD $56694$ - $56713$) and B (MJD $56967$ - $56980$) for four different energy bands. {\bf (a)}: LCs of Flare-A in four energy bands i.e. E1:$0.1-300$ GeV (topmost panel), E2:$0.1-0.3$ GeV ($2$nd panel), E3:$0.3-1$ GeV ($3$rd panel) and $1-300$ GeV (bottommost). {\bf (b)}: Similar LCS for Flare-B. The red curve represents the best-fitted model.}
\label{fig:flare_fitting}
\end{figure*}
\subsection{Flare Fitting}
To explore the subflare structures, energy-resolved Fermi-LAT LC with the shortest possible time bin is required. We divided the Fermi energy range ($0.1-300$ GeV) into three consecutive bands i.e. E1: $0.1-0.3$ GeV, E2: $0.3-1$ GeV and E3: $1-300$ GeV and generated $12$hr binned (shortest possible time bin as allowed by photon statistics) LC for each energy band. Due to short time binning and fixed criteria of minimum $3\sigma$ significance, a significant number of time bins provided an upper limit instead of flux estimate. It constrained us to consider only a short portion of the entire flare, and we have chosen the same epochs as in the case of the flux-index hysteresis curve. We fitted each LC with a function which is the sum of a time-dependent flux profile and a constant baseline flux ($C$). This relatively simple decomposition is very useful in probing short flux variations \citep{Villata1999A&A}. It also provides valuable information about the acceleration and cooling timescales of parent particles. Due to the asymmetric nature of flares, the time-dependent flux profile is chosen as,
\begin{equation}
\label{eq:asy_exp}
    F(t) = 2F_0(exp^{(t_0 - t)/T_R} + exp^{(t - t_0)/T_D})^{-1}
\end{equation}

where $F(t)$ represents flux at time $t$, $F_0$ is the local peak flux and $t_0$ is the corresponding time, $T_R$ and $T_D$ represent the rise and decay timescales of the corresponding subflares respectively \citep{Abdo2010cApJ}. By fitting a constant to the selected portion of $12$-hour binned LC in each energy band, the baseline flux ($C$) was derived. In the subsequent fitting, $C$ was allowed to vary over a short-range around its best-fitted value. $t_0$ was initially determined from the data set and allowed to vary over one time bin. Individual fitting of each subflare provided the initial values of $F_0$, $T_R$, and $T_D$. Final fitted values were obtained from the combined fitting of all subflares. 

The modelled LCs for all energy bands of both flare states are depicted in \autoref{fig:flare_fitting}.           \autoref{fig:flare_fitting}-a and \autoref{fig:flare_fitting}-b represent the modelled LCs during Flare-A and B respectively. The uppermost panels in both represent the combined ($0.1-300$ GeV) 12-hour binned Fermi-LAT LC over the selected flare region (i.e. MJD $56694$ -- $56713$ and MJD $56967$ -- $56980$ for Flare-A and B respectively). The combined Fermi LC is presented here for comparison with individual energy bands. The other three panels in  \autoref{fig:flare_fitting}-a and \autoref{fig:flare_fitting}-b show the fitted LC for each energy band in ascending order. The results of the fit and corresponding errors are given in \autoref{tab:flare_fit_par}. The subflares are numbered in chronological order. We have also estimated the asymmetry parameter (A) for each sub-flare which is defined as, 
\begin{equation}
    A = \frac{|(T_R-T_D)|}{(T_R+T_D)}
\end{equation}

where $T_R$ and $T_D$ represent the rise and decay time of the corresponding subflare. From \autoref{fig:flare_fitting}, it is clear that Flare-B is the combination of two prominent subflares which are seen in all three bands. On the contrary, Flare-A is the result of a single subflare except in the E1 energy band. Another interesting feature of Flare-B is that the relative contribution of subflare-1 is gradually decreasing, whereas it is increasing with an increase in energy for subflare-2. These features are explored later in the paper. The close examination of \autoref{tab:flare_fit_par} shows that, in the case of Flare-B, the rise and decay times are well correlated for the first two bands and change abruptly in the 3rd one.
\begin{table*}
	\centering
	\caption{Flare characterization: Best fit model parameters and corresponding errors}
	\label{tab:flare_fit_par}
	\scalebox{1.0}{
	\begin{tabular}{lccccccccc}
		\hline
		\hline
		Flare & Energy Band & subflare & $C$ & $t_0$ & $F_0$ &$T_R$ & $T_D$ & $\chi^{2}/\text{(dof)}$ & A \\
	state	& (GeV) & & ($10^{-7} \text{ph cm}^{-2} \text{s}^{-1}$) & (MJD) & ($10^{-7} \text{ph cm}^{-2} \text{s}^{-1}$) & (days) & (days) & & \\
		\hline
		A & $0.1 - 0.3$ & subflare-1 & $5.19 \pm 0.92$ & $56699.25 \pm 0.5$ & $8.24 \pm 2.09$ & $1.19 \pm 0.42$ & $0.17 \pm 0.12$ & $0.88$ & $1.06$ \\
		 &  & subflare-2 & $5.19 \pm 0.92$ & $56702.25 \pm 0.5$ & $18.57 \pm 0.67$ & $0.31 \pm 0.19$ & $0.82 \pm 0.23$ & & $0.42$ \\
	   \hline
         & $0.3 - 1$ & subflare-2 & $1.94 \pm 0.31$ & $56702.75 \pm 0.5$ & $7.86 \pm 0.66$ & $0.7 \pm 0.14$ & $0.34 \pm 0.16$ & $0.79$ & $0.37$ \\
       \hline
         & $1 - 300$ & subflare-2 & $0.55 \pm 0.15$ & $56702.25 \pm 0.5$ & $1.73 \pm 0.06$ & $0.6 \pm 0.22$ & $1.27 \pm 0.05$ & $1.01$ & $0.08$ \\
        \hline
        \hline
       B & $0.1 - 0.3$ & subflare-1 & $10.27 \pm 1.09$ & $56973.25 \pm 0.5$ & $46.79 \pm 7.3$ & $0.32 \pm 0.1$ & $0.56 \pm 0.1$ & $0.87$ & $0.32$ \\
         &  & subflare-2 & $10.27 \pm 1.09$ & $56976.25 \pm 0.5$ & $25.1 \pm 3.4$ & $0.31 \pm 0.15$ & $0.6 \pm 0.16$ & & $0.35$ \\
        \hline
        & $0.3 - 1$ & subflare-1 & $2.77 \pm 0.36$ & $56973.25 \pm 0.5$ & $9.33 \pm 3.21$ & $0.34 \pm 0.13$ & $0.57 \pm 0.14$ & $1.15$ & $0.29$ \\
         &  & subflare-2 & $2.77 \pm 0.36$ & $56976.25 \pm 0.5$ & $7.21 \pm 1.38$ & $0.33 \pm 0.17$ & $0.69 \pm 0.4$ & & $0.35$  \\
         \hline
         & $1 - 300$ & subflare-1 & $0.64 \pm 0.17$ & $56972.75 \pm 0.5$ & $1.53 \pm 0.92$ & $0.29 \pm 0.33$ & $0.73 \pm 0.29$ & $0.22$ & $0.43$ \\
         &  & subflare-2 & $0.64 \pm 0.17$ & $56976.75 \pm 0.5$ & $2.84 \pm 1.25$ & $0.45 \pm 0.17$ & $0.2 \pm 0.12$ & & $0.14$ \\
		\hline
	\end{tabular}
	}
\end{table*}
\subsection{Hardness Ratio}
\begin{figure*}
    \centering
    \includegraphics[trim={3.7cm 0 5cm 0},clip, width=0.498\linewidth, height=7cm]{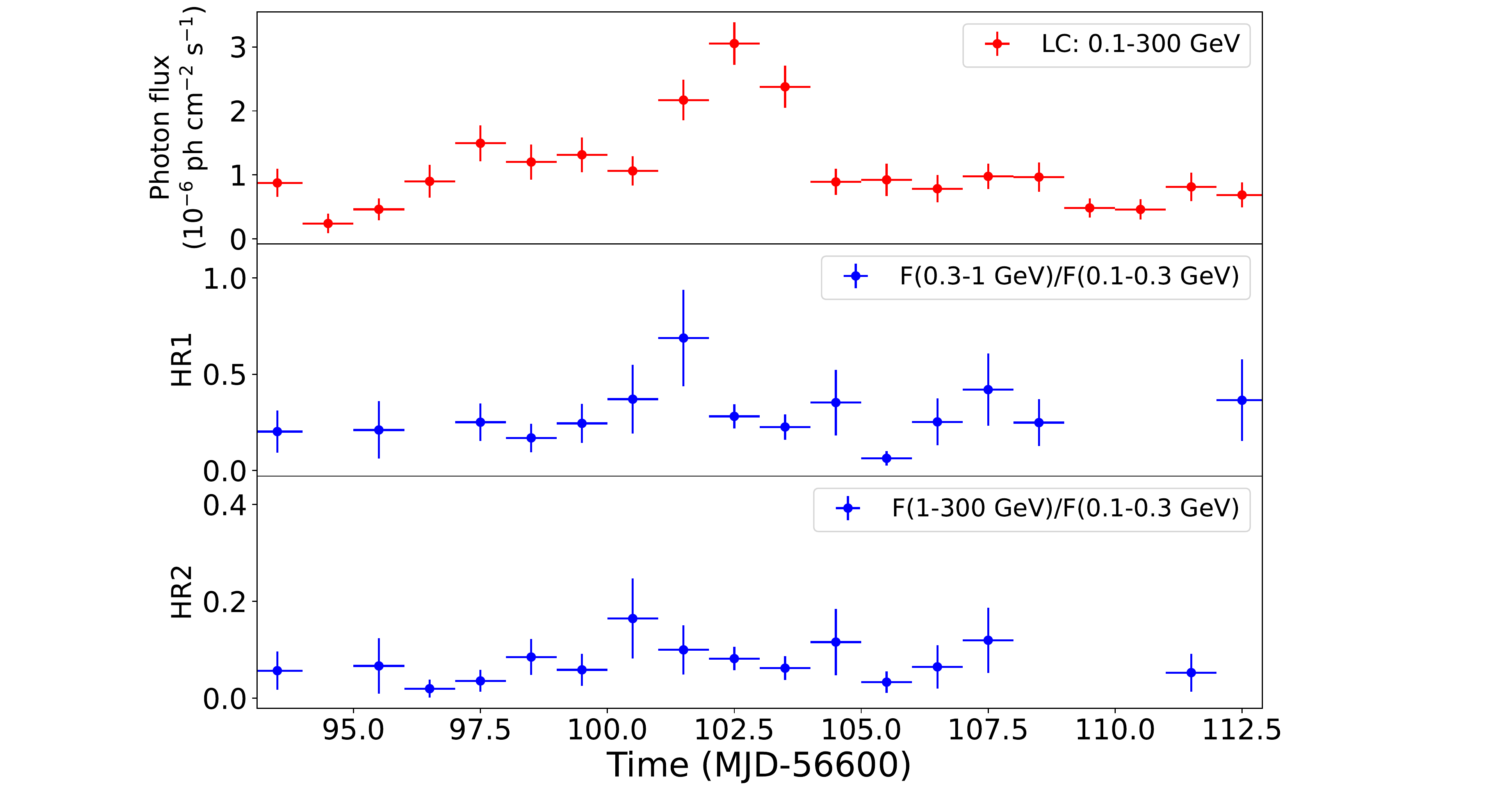}
	\includegraphics[trim={3.7cm 0 5cm 0},clip, width=0.498\linewidth, height=7cm]{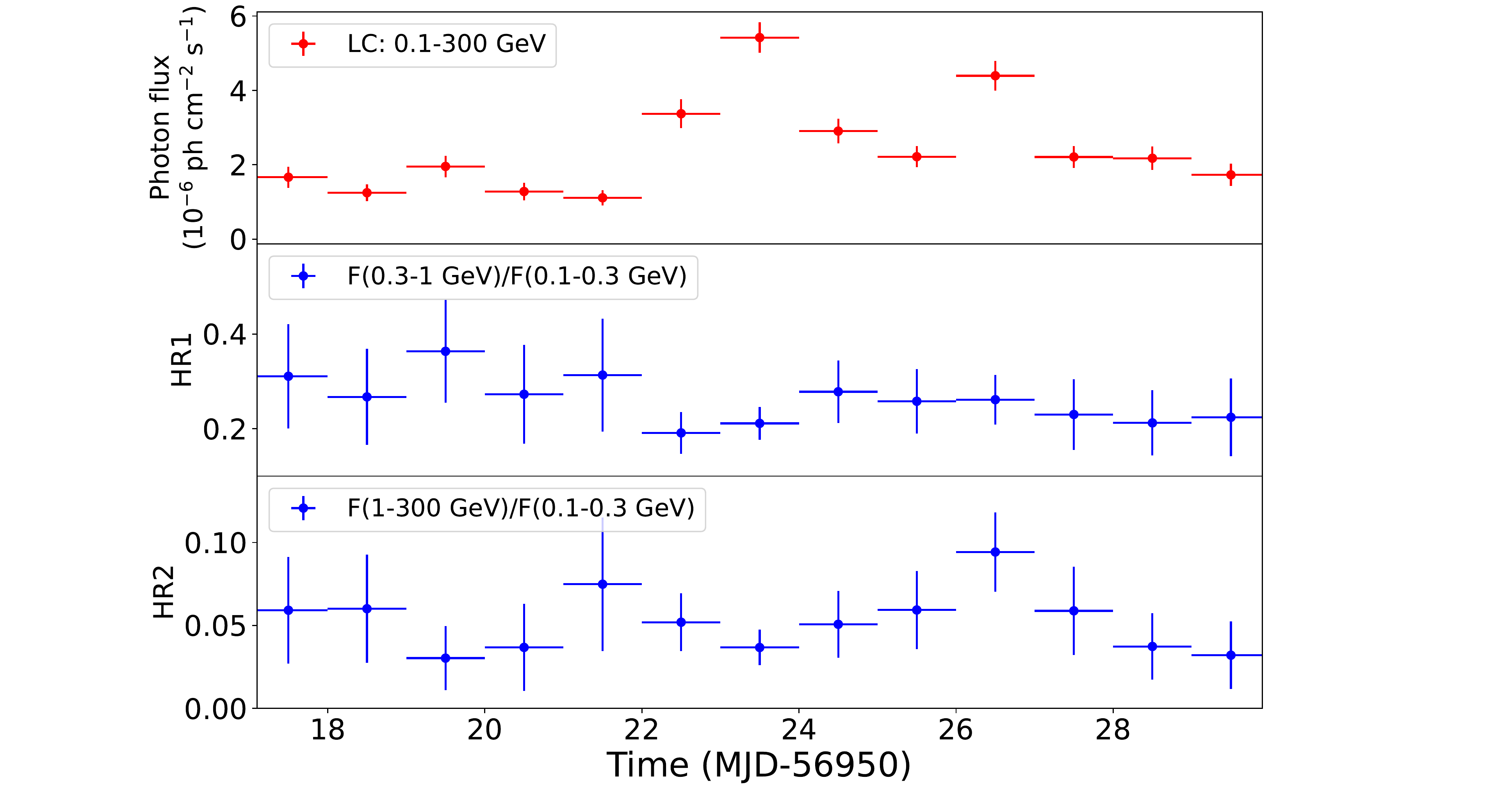}
	\caption{
    Daily binned Fermi LC ($0.1-300$ GeV) and Hardness ratios on a daily timescale during Flare-A (left panel) and B (right panel). HR1: $F_{0.3-1 \text{ GeV}}/F_{0.1-0.3 \text{ GeV}}$ and HR2: $F_{1-300 \text{ GeV}}/F_{0.1-0.3 \text{ GeV}}$. The middle and bottom panels represent the evolution of HR1 and HR2, respectively.}
	\label{fig:Hardness_ratio}
\end{figure*}

In this work, we have studied the hardness ratio (HR) between different energy bands of the GeV regime during both the flare states. The HRs are defined to explore the relative contributions of individual energy bands. We have used two hardness ratios in this work, HR1 and HR2. HR1 is the ratio of the photon flux in the E2 band to that in the E1 band. Similarly, HR2 is the ratio of the photon flux in the E3 band to that in the E1 Band. Daily binned Fermi LCs are used for the calculation of the hardness ratio. The left and right panels of \autoref{fig:Hardness_ratio} represent daily binned $0.1-300$ GeV LC along with the hardness ratio during Flare-A and B, respectively. The error bars of HRs are derived from individual flux errors with simple error propagation formula. The hardness ratio provides a unique evolution trend at rising and decaying portions that are different from flare evolution. In Flare-A, the plateau portion before the peak flux in daily binned LC is absent in both HR1 and HR2. The hardening of HR1 and HR2 are stopped well before the peak flux position, and HR2 shows a local minimum corresponds to that peak. In the case of flare-B, HR1 shows nearly constant behaviour throughout the flare period. On the contrary, HR2 represents a clear minimum and a prominent peak at the position corresponding to subflare-1 and subflare-2, respectively. 
\subsection{Flare State Modelling}
The modelling of SED over a broad energy range is a helpful tool to probe the flaring mechanism. It can provide a detailed insight into different acceleration mechanisms. The flare states' SED modelling of PKS B1222+216 with the leptonic scenario and its evolution are discussed here.

\subsubsection{Model Description}
In this work, single-zone leptonic model based on the stochastic or shock acceleration mechanism \citep{Ghisellini2010MNRAS,Tramacere2011ApJ,Ghisellini2009MNRAS} has been considered and implemented using a numerical code \lq JETSET\rq \citep{Massaro2006A&A,Tramacere2009A&A,Tramacere2011ApJ,Tramacere2020ascl}. The model assumes the presence of background plasma consisting of relativistic electrons and positrons in the jet. These leptons are accelerated to an ultra-relativistic regime via a stochastic or shock acceleration mechanism inside the jet.
For simplicity, a spherically symmetric emission region, travelling with a bulk Lorentz factor of $\Gamma$, is considered inside the jet. Due to the relativistic motion, the observed emission is Doppler boosted by a factor $\delta = 1/[\Gamma(1-\beta \cos{\theta})]$, where $\theta$ is the angle between the jet axis and observer's line of sight. In this work, we have assumed $\delta = \Gamma$.
The model also assumes that the acceleration region is filled with homogeneous lepton population, and its energy distribution can be described by a broken power-law model as given by:
\begin{equation}
    n(\gamma) =
    \begin{cases}
    k \gamma^{-p_1} & \text{ when } \gamma_{min} \leq \gamma \leq \gamma_{b} \\
    k \gamma_b^{p_2 - p_1}\gamma^{-p_2} & \text{ when } \gamma_b < \gamma \leq \gamma_{max}
    \end{cases}
\end{equation}
and
\begin{equation}
    \int_{\gamma_{min}}^{\gamma_{max}}n(\gamma)d\gamma=N 
\end{equation}

where $N$ represents the total number of emitting particles per unit volume (\#/$\text{cm}^3$) and $k$ is the normalization constant. $\gamma_{min}$ and $\gamma_{max}$ represent the lowest and the highest available electron energy in the population. $\gamma_b$ represents the break-energy where the slope of energy distribution changes from $p_1$ to $p_2$.

Regarding the characterization of the emission zone, the model invokes two physical parameters i.e. its radius ($R$) and distance from the central engine ($d$). The size of the emission region ($R$) is constrained by the light travel time argument, which is given by,
\begin{equation}
    R \leq \frac{c\delta t_{var}}{2(1+z)}
\end{equation}
where $t_{var}$ and $z$  respectively represent the flux variability time scale and cosmological red shift. For the estimation of flux variability time scale, we have used the daily binned Fermi LC data as shown in \autoref{fig:dailly-LC}. All the consecutive pair of flux points (excluding upper limits) of Fermi LC have been chosen and scanned for the fastest variability timescale using the following analytical formula:
\begin{center}
\begin{equation}
    F(t_2) = F(t_1) 2^{(t_2 - t_1)/t_{var}}
\end{equation}
\end{center}
where $F(t_1)$  and $F(t_2)$ is the photon flux at initial time $t_1$ and later time $t_2$ respectively. To acquire confidence level, we also estimated the significance corresponding to each pair. The final variability time scale is obtained by averaging all available data that meet the minimum significance criteria of $3\sigma$. These values are given in \autoref{tab:basic_par}.
\begin{table}
	\centering
	\caption{Basic model parameters of PKS B1222+216}
	\label{tab:basic_par}
	\scalebox{0.85}{
	\begin{threeparttable}
	\begin{tabular}{lccccccc} 
		\hline
		Flare & $\text{z}$ & $\theta$ \tnote{1} & $\Gamma$ & $\delta$ \tnote{1} & $M_{BH}$ \tnote{2} & $L_D$ \tnote{2} & $t_{var}(\geq 3\sigma)$\\
		 state & & (deg) & & & $(M_{\odot})$ & $(10^{46}\text{ erg/s})$ & (days) \\
		\hline
		\hline
		A & 0.432 & 2.5 & 23.0 & 23.0 & $6.0\times10^8$ & 3.5 & 0.83\\
		B &  &  &  &  &  & & 0.99\\
		\hline
	\end{tabular}
	\end{threeparttable}
	}
	\begin{tablenotes}
		\item[1] \citep{Kushwaha2014ApJ}; \item[2] \citep{Farina2012MNRAS};
	\end{tablenotes}
\end{table}

After being accelerated, the lepton distribution gets cooled by non-thermal photon emission mechanisms. The interaction with the entangled magnetic field inside the emission region produces synchrotron emission and generates the low energy hump in the observed SED. 
On the other hand, the up-scattering of the low energy seed photons produce the high-energy hump as a combination of different IC components (EC-BLR, EC-DT and SSC).

The distance of BLR ($R_{BLR}$) and dusty torus region ($R_{DT}$) from the central engine are estimated using reverberation mapping technique \citep{Ghisellini2010MNRAS} and given by,
\begin{equation}
    R_{BLR} = 10^{17} \sqrt{L_D/10^{45}} \text{ cm}
\end{equation}
\begin{equation}
    R_{DT} = 2.5\times10^{18} \sqrt{L_D/10^{45}} \text{ cm}
\end{equation}
where $L_D$ represents the disc luminosity in the unit of erg/sec. In this model, the BLR is considered a thin spherical shell of a typical width of $0.01$ pc made of ionised gas. Thus the inner and outer radius of BLR region are estimated as $R_{BLR}^{in}=(R_{BLR}-0.005\text{ pc})$ and $R_{BLR}^{out}=(R_{BLR}+0.005\text{ pc})$. The temperature of the dusty torus region is kept fixed at $1000$ K.
The thermal emission of the accretion disc is modelled as the multi-temperature blackbody radiation where the temperature at a particular position depends on its distance ($r$) from the central engine and varies as,
\begin{equation}
    T^4(r) = \frac{3R_S L_D}{16\epsilon \upi \sigma r^3}\left(1-\sqrt{\frac{3R_S}{r}}\right) \text{ K}
\end{equation}
where $R_S$ is the Schwarzschild radius and $\epsilon$ represents the accretion efficiency. The radial extent of accretion disc is assumed from $3R_S$ to $500R_S$. $\sigma$ is the Stefan-Boltzman constant. $\tau_{BLR}$ and $\tau_{DT}$ represent the fraction of disc energy absorbed and emitted by BLR and dusty torus region, respectively. This model first calculates the luminosity in the moving frame attached with the emission region and finally transforms into the observed rest frame with the help of observed red shift information. A $\Lambda\text{CDM}$ flat cosmological model with $H_{0} = 67.11 \text{ km}\text{s}^{-1}\text{Mpc}^{-1}$, $\Omega_{m} = 0.3175$ and $\Omega_{\Lambda} = 0.3175$ is used in this work \citep{Planck2014A&A}.

\subsubsection{Flare-A (MJD 56680--56748)}
\begin{figure*}
    \centering
    \includegraphics[trim={0.5cm 0 0 0},clip, width=1\linewidth]{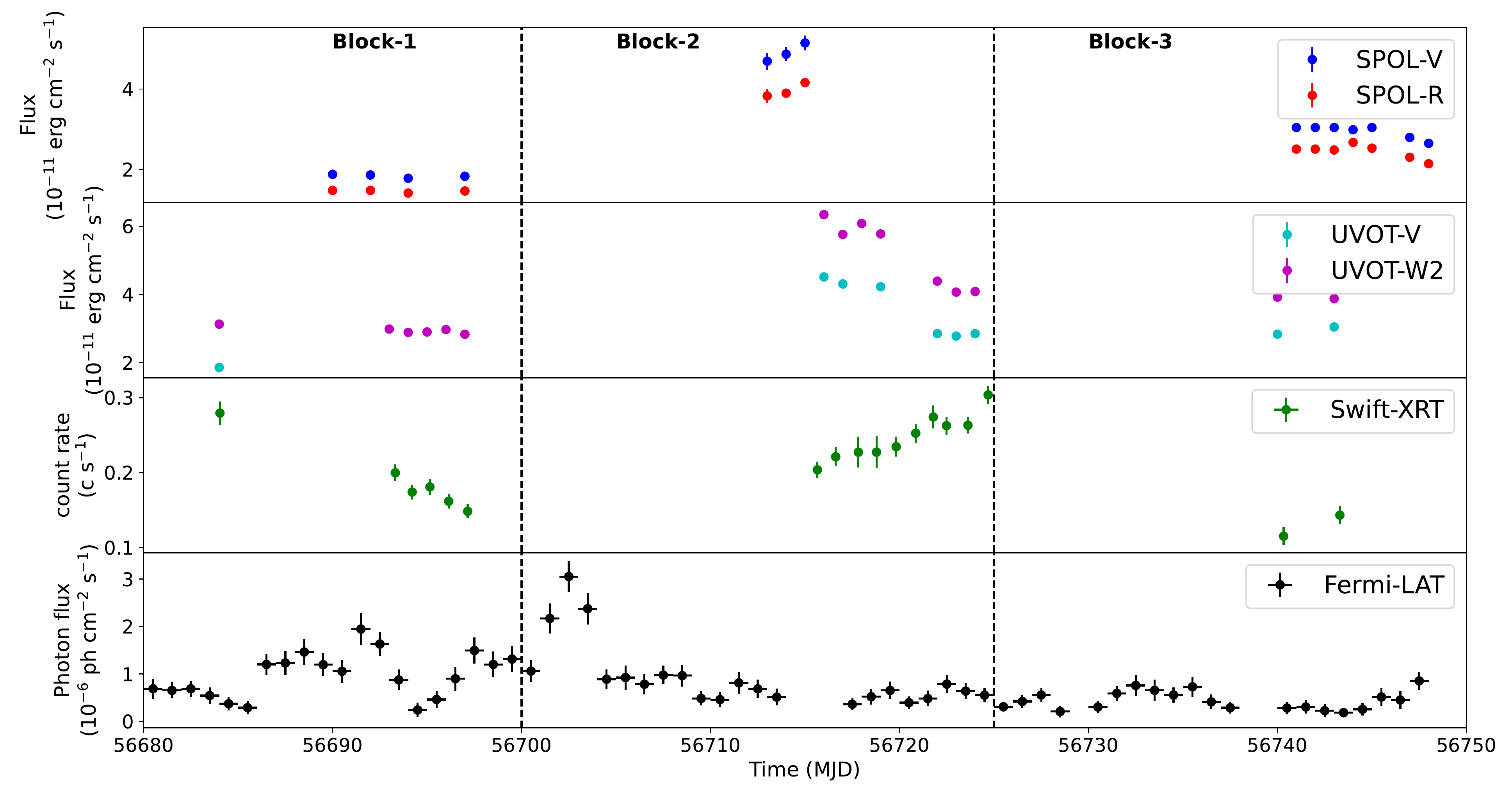}
	\caption{The four panels from top to bottom order represent light curves of PKS B1222+216 corresponds to optical, UV, soft X-ray, and MeV-GeV bands during Flare-A. The V and R filter data of the SPOL-CCD instrument use as optical data. The \textit{Swift}-UVOT data taken with V and W2 filters avail as UV data. The \textit{Swift}-XRT (0.3 - 8.0 keV) and \textit{Fermi}-LAT (0.1 - 300 GeV) data respectively used as soft X-ray and MeV-GeV data.}
	\label{fig:LC_comparison}
\end{figure*}

SED of Flare-A consists of data from all the instruments mentioned in section 2. Corresponding LCs are shown in figure~\autoref{fig:LC_comparison}. There is an indication of time lags in various wavebands relative to $\gamma-$ray LC with the maximum apparent lag in the X-ray band.          
It should be noted that sampling in various wavebands is rather sparse. Between MJD $56710-56720$, a huge flux enhancement (2--3 times the previous flux) has been observed in both optical and all available filters of UV bands which decayed in later periods. In the X-ray band, a decreasing trend in count rate just before the GeV peak flux (before MJD $56703$) has been converted into an increasing pattern during the period of MJD $56714-56725$. In these multi-band LCs, though the highest flux states in different bands arise at different times, they are distributed within a span of MJD $56700 \text{ to }56725$. For a detailed study of Flare-A, we have divided the whole flare period into three blocks of nearly equal length, and broadband modelling has been carried out individually for each block. This partition in blocks is also helpful in avoiding the averaging of low and high flux states. The individual block details are given in \autoref{tab:Block_details}. The length of the middle block (Block-2) is chosen carefully to contain the peak flux states from all bands and, therefore, the first and third blocks represent SEDs just before and after the maximum flaring activity. This block division also allows us to visualize the gradual change in the underlying particle spectrum which governs the flare.
\begin{table}
	\centering
	\caption{Details of the block distribution of Flare-A}
	\label{tab:Block_details}
	\begin{tabular}{|c|c|c|c|} 
		\hline
		  & Block-1 & Block-2 & Block-3\\
		\hline
		Time Range (MJD) & 56680 - 56700 & 56700 - 56725 & 56725 - 56748\\
		\hline
		Duration (days) & 20 & 25 & 23\\
		\hline
	\end{tabular}
\end{table}

\begin{figure*}
    \centering
	\includegraphics[width=0.49\linewidth]{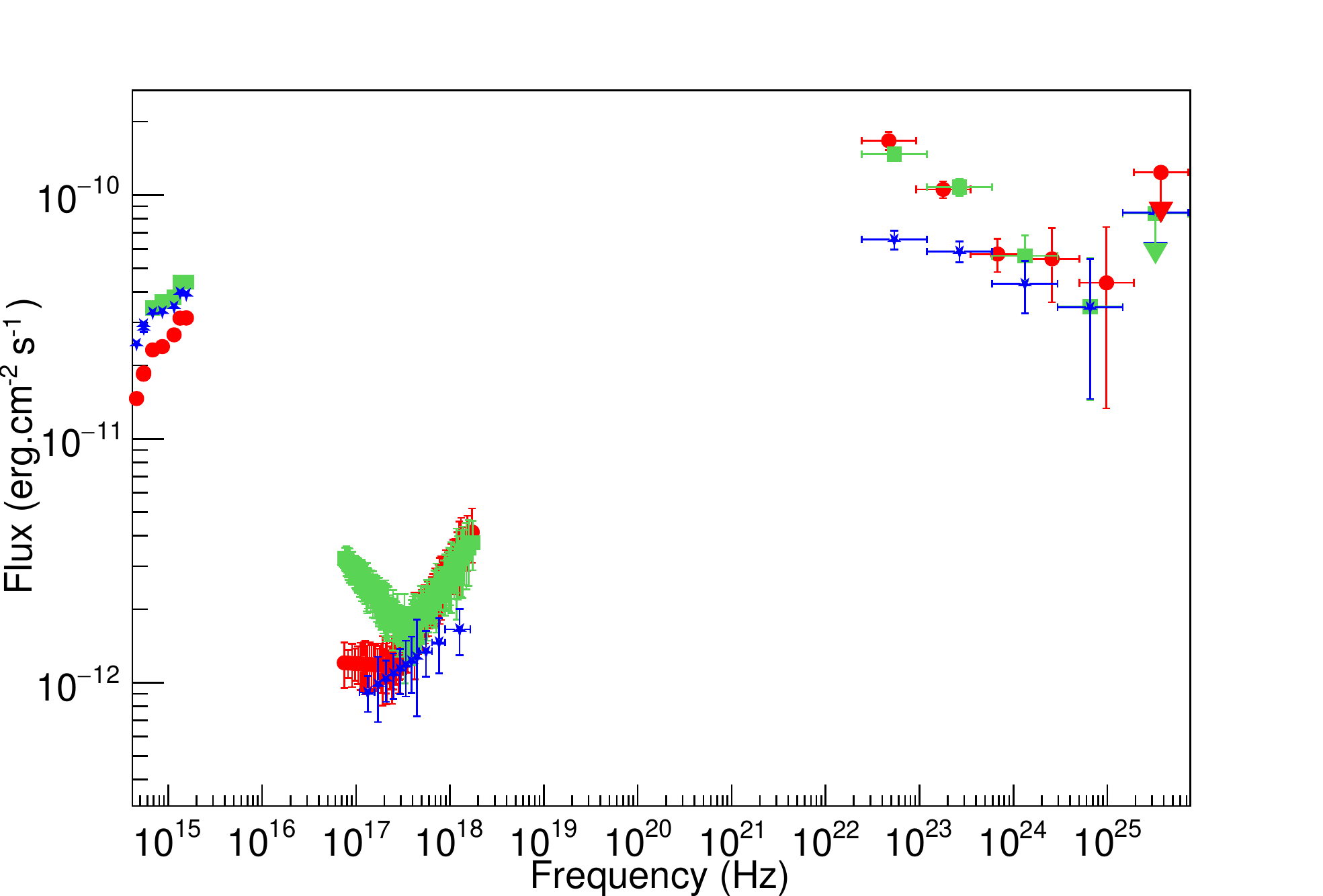}
	\includegraphics[width=0.49\linewidth]{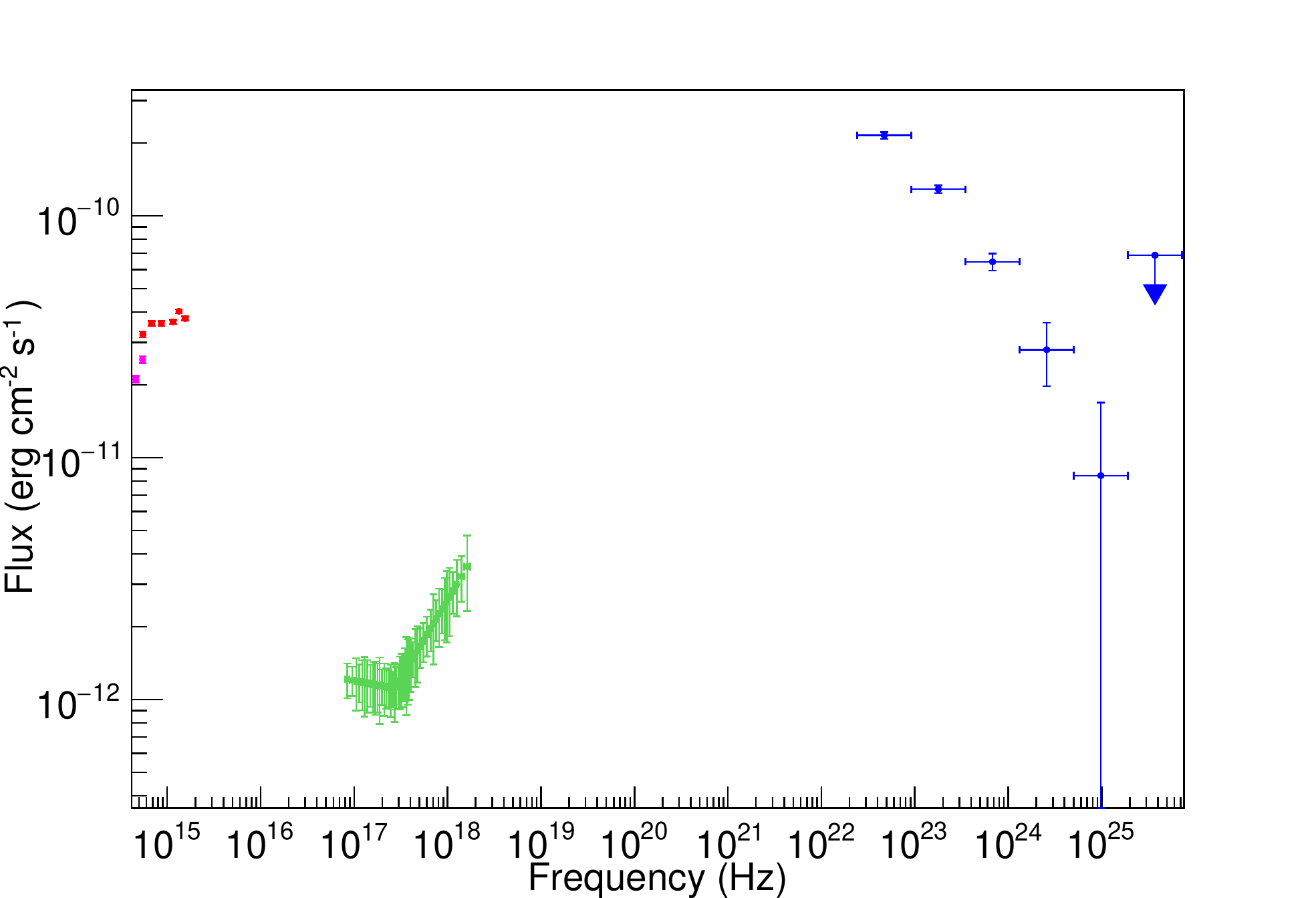}
	\caption{
    Observed broadband SED of Flare-A and B. Left panel shows SEDs corresponding to all blocks of Flare-A. SED of Block-1, Block-2 and Block-3 are represented by red, green and blue data points, respectively. The right panel represents SED of flare-B where different colours represent different data sets (Pink: SPOL-CCD, Red: \textit{Swift}-UVOT, Green: \textit{Swift}-XRT and Blue: Fermi-LAT).}
	\label{fig:Obs_SED}
\end{figure*}

The broadband data from different instruments are processed using the dedicated software tools as mentioned in \autoref{sec:obs}. In the optical band, a total of $14$ observations are available from the SPOL-CCD during Flare-A.
The distribution of these observations among different blocks and average flux for both V and R band filters after the reddening correction are given in \autoref{tab:opt_details}.
The \textit{Swift}-UVOT and XRT instruments provide a total of $18$ observations in UV and X-ray bands during Flare-A.
\autoref{tab:uvot_details} shows the distribution of these observations among blocks and also provides the average optical and UV fluxes for all six filters after the corresponding reddening corrections.
Details of results from X-ray data analysis i.e. spectral model, fitted parameter values, and goodness of fit are given in \autoref{tab:xrt_details}. The fluxes are corrected for the galactic soft X-ray absorption.
Details of Flare-A GeV data analysis are given in \autoref{tab:lat_details}. A log-parabola model with fixed break-energy ($E_b$) is used for spectral modelling of all the blocks.
The left panel of \autoref{fig:Obs_SED} shows block-wise broadband SED data of Flare-A. The optical-UV band shows an increase of flux in Block-2 and Block-3 compared to Block-1 which is consistent with the SPOL and UVOT LCs of \autoref{fig:LC_comparison}. One of the most interesting features is the dramatic change in the X-ray spectral shape during Flare-A. 
Starting with a broken power-law spectral shape with low break-energy in Block-1, a high-energy shift in both break-energy and corresponding spectral shape occurs in Block-2. Finally, it returns to a power-law form in Block-3. This abrupt change in X-ray spectral shape justifies the block distribution over the flare period. In the $\gamma-$ray band, starting with a steeper shape in Block-1 and Block-2, the spectrum gets flattened in Block-3.

In SED modelling, each available data provide some constraints on the parameter space. Along with the analyzed data, we have used the archival data from SSDC\footnote{SSDC (ASI): \href{https://www.ssdc.asi.it}{https://www.ssdc.asi.it}}. In the case of blazars, some general constraints are as follows:

i)  Due to synchrotron self-absorption, the synchrotron emission component shows a sharp decline below $\sim10^{12}$ Hz. The observed radio flux points are above the predicted curve and indicate
emission associated with the extended region of jet emission where the self-absorption cross-section is significantly less. Thus radio data provides only a little constraint on the low energy synchrotron spectrum as an upper limit.


ii) Optical-UV data forms a little hump in the observed SED of most FSRQs. It is interpreted as the combination of synchrotron and direct thermal emission from the disk. The fitting of the optical data provides a constraint on the magnetic field ($B$), particle density ($N$), and the first index of particle spectrum $(p_1)$.

iii) Simultaneous mm-IR observations provide a good constraint on the location of the peak synchrotron emission. Due to the lack of IR observations, it is not possible to infer the first hump (synchrotron) peak position beforehand.

iv) The shape of X-ray (particularly high energy tail) and $\gamma-$ray data suggest that they are part of the single hump, and both the X-ray and $\gamma-$ray data interpret as the combination of both EC and SSC emission. The slope of the $\gamma-$ray spectrum constrains the high energy slope ($p_2$) of the particle distribution, whereas the position of the lowest energy flux point of Fermi-LAT data provides an upper limit for the break-energy ($E_b$).


v) Disk luminosity ($L_D$), the mass of SMBH ($M_{BH}$), bulk Lorentz factor ($\Gamma$), and viewing angle ($\theta$) kept frozen to their literature values.

vi) Due to the sparse nature of the data, non-frozen parameters are fitted by eye.

In this model, we have considered that the blob is moving within the jet.  Therefore its distance from the central engine is changing with time. In Block-1 modelling, blob position ($d$) was obtained from SED fitting, whereas for the other two Blocks, it is partially constrained by the actual distance traversed by the blob within those time intervals. In the rest frame of SMBH, the distance covered by the blob provides an upper limit on the change in $d$.
\begin{table*}
	\centering
	\caption{Details of optical flux measurements from SPOL-CCD observations. Effective wavelengths ($\lambda_{\text{eff}}$) and corresponding reddening corrections ($A_{\lambda}$) are mentioned for both V and R filters. }
	\label{tab:opt_details}
	\begin{tabular}{ccccccccc} 
		\hline
		 Flare state & Block & No. of Obs & \multicolumn{3}{c}{V band} & \multicolumn{3}{c}{R band}\\
		 \cline{4-6}   \cline{7-9}
		  &  &  & $\lambda_{\text{eff}}$ & $A_{\lambda}$ & Average flux & $\lambda_{\text{eff}}$ & $A_{\lambda}$ & Average flux\\
		 & & & ({$\mathring{\text{A}}$}) & & ($10^{-11}$ erg $\text{cm}^{-2}\text{s}^{-1}$) & ({$\mathring{\text{A}}$}) & & ($10^{-11}$ erg $\text{cm}^{-2}\text{s}^{-1}$)\\
		\hline
		\hline
		A & Block-1 & 4 & $5517$ & $0.062$ & $1.84 \pm 0.07$ & $6520$ & $0.049$ & $1.46 \pm 0.04$\\
		 & Block-2 & 3 & & & $4.92 \pm 0.19$ & & & $3.96 \pm 0.13$\\
		 & Block-3 & 7 &  & & $2.93 \pm 0.11$ & & & $2.45 \pm 0.67$\\
		\hline
		B &  & 8 & & & $2.55 \pm 0.88$ & & & $2.11 \pm 0.66$\\
		\hline
	\end{tabular}
\end{table*}
\begin{table*}
	\centering
	\caption{Result of \textit{Swift}-UVOT data analysis. Average fluxes are given in the unit of $10^{-11}$ erg $\text{cm}^{-2}\text{s}^{-1}$.}
	\label{tab:uvot_details}
	\scalebox{0.98}{
	\begin{tabular}{lcccccccc} 
		\hline
		 Flare & Block & Obs ID & V & B & U & W1 & M2 & W2\\
		\hline
		\hline
		A & Block-1 & 00036382033 - 00036382034 & $1.86 \pm 0.07$ & $2.31 \pm 0.06$ & $2.39 \pm 0.07$ & $2.67 \pm 0.06$ & $3.12 \pm 0.07$ & $3.13 \pm 0.06$\\
		& & 00036382036 - 00036382039 & & & & & & \\
		\cline{2-9}
		 & Block-2 & 00036382040 & $2.85 \pm 0.08$ & $3.44 \pm 0.08$ & $3.66 \pm 0.09$ & $3.81 \pm 0.08$ & $4.39 \pm 0.09$ & $4.39 \pm 0.08$\\
		 & & 00036382045 - 00036382053 & & & & & & \\
		 \cline{2-9}
		 & Block-3 & 00036382054 - 00036382055 & $2.84 \pm 0.10$ & $3.3 \pm 0.09$ & $3.32 \pm 0.09$ & $3.49 \pm 0.08$ & $3.98 \pm 0.1$ & $3.92 \pm 0.08$\\
		\hline
		B &  & 00036382058 & $3.24 \pm 0.1$ & $3.59 \pm 0.08$ & $3.59 \pm 0.09$ & $3.64 \pm 0.08$ & $4.03 \pm 0.08$ & $3.76 \pm 0.07$\\
		& & 00036382060 - 00036382061 & & & & & & \\
		\hline
	\end{tabular}
	}
\end{table*}
\begin{table*}
	\centering
	\caption{Result of \textit{Swift}-XRT data analysis. Observation IDs are same as given in \autoref{tab:uvot_details}. The best fitted spectral model and value of its parameters like index ($\Gamma_1$) and normalization constant ($K$) for power law and first/second index ($P_1$/$P_2$), break-energy ($\text{E}_b$) and normalization constant ($K$) for broken power law are given. The integrated flux values over the 0.3 - 8.0 keV band are also given. We use a hydrogen column density of $\text{n}_\text{H} = 1.72 \times 10^{20} \text{cm}^{-2}$ \citep{HIsurvey2016A&A}.}
	\label{tab:xrt_details}
	\scalebox{0.91}{
	\begin{tabular}{lccccccccc} 
		\hline
		 Flare & Block & Model & $\Gamma_1$ & $P_1$ & $P_2$ & $E_b$ & $K$ & $\chi_{r}^2$ & $\text{F}_{0.3-8.0\text{ keV}}$ \\
		 state & & & & & & (keV) & ($10^{-4}\text{ ph}\text{ cm}^{-2}\text{s}^{-1}\text{keV}^{-1}$) & & ($10^{-12}\text{ erg}\text{  cm}^{-2}\text{s}^{-1}$) \\
		\hline
		\hline
		A & Block-1 & Broken Power Law & & $2.02 \pm 0.17$ & $1.36 \pm 0.06$ & $0.98 \pm 0.17$ & $7.30 \pm 0.71$ & $0.85$ & 6.58\\
		 & Block-2 & Broken Power Law & & $2.46 \pm 0.06$ & $1.46 \pm 0.07$ & $1.46 \pm 0.11$ & $11.84 \pm 0.34$ & 1.14 & 8.14\\
		 & Block-3 & Power Law & $1.73 \pm 0.11$ & & & & $6.74 \pm 0.55$ & $0.89$ & 4.12\\
		 \hline
	   B &  & Broken Power Law & & $2.07 \pm 0.18$ & $1.35 \pm 0.09$ & $1.13 \pm 0.21$ & $6.99 \pm 0.66$ & $0.91$ & 5.95\\
		 \hline
	\end{tabular}
	}
\end{table*}
\begin{table*}
	\centering
	\caption{Result of \textit{Fermi}-LAT spectral analysis. The normalization constant ($N_0$), spectral indices ($\alpha$ and $\beta$), integral photon flux ($F_{0.1-300\text{ GeV}}$) and corresponding test statistic (TS) for unbinned analysis are given.}
	\label{tab:lat_details}
	\scalebox{1.0}{
	\begin{tabular}{lccccccc} 
		\hline
		 Flare  & Block & $\alpha$ & $\beta$ & $\text{E}_\text{b}$ & $\text{N}_0$ & $\text{F}_{0.1-300 \text{ GeV}}$ & TS \\
		 state & & & & (MeV) & ($10^{-10}\text{ ph}\text{ cm}^{-2}\text{s}^{-1}\text{MeV}^{-1}$) & ($10^{-7}\text{ ph}\text{ cm}^{-2}\text{s}^{-1}$) & \\
		\hline
		\hline
		A & Block-1 & $2.36 \pm 0.05$ & $0.04 \pm 0.02$ & $393.682$ & $5.22 \pm 0.24$ & $9.74 \pm 0.51$ & $1733.07$\\
	    & Block-2 & $2.22 \pm 0.02$ & $0.03 \pm 0.01$ &  & $5.34 \pm 0.11$ & $8.94 \pm 0.24$ & $2012.42$\\
	    & Block-3 & $2.09 \pm 0.09$ & $0.01 \pm 0.04$ &  & $2.54 \pm 0.2$ & $4.01 \pm 0.36$ & $834.47$\\
	    \hline
	   B &  & $2.41 \pm 0.037$ & $0.03 \pm 0.02$ &  & $6.65 \pm 0.21$ & $12.43 \pm 0.35$ & $6117.95$\\
	   \hline
	\end{tabular}
	}
\end{table*}

\begin{figure*}
\centering
\begin{tabular}{cc}
\includegraphics[width=0.5\textwidth]{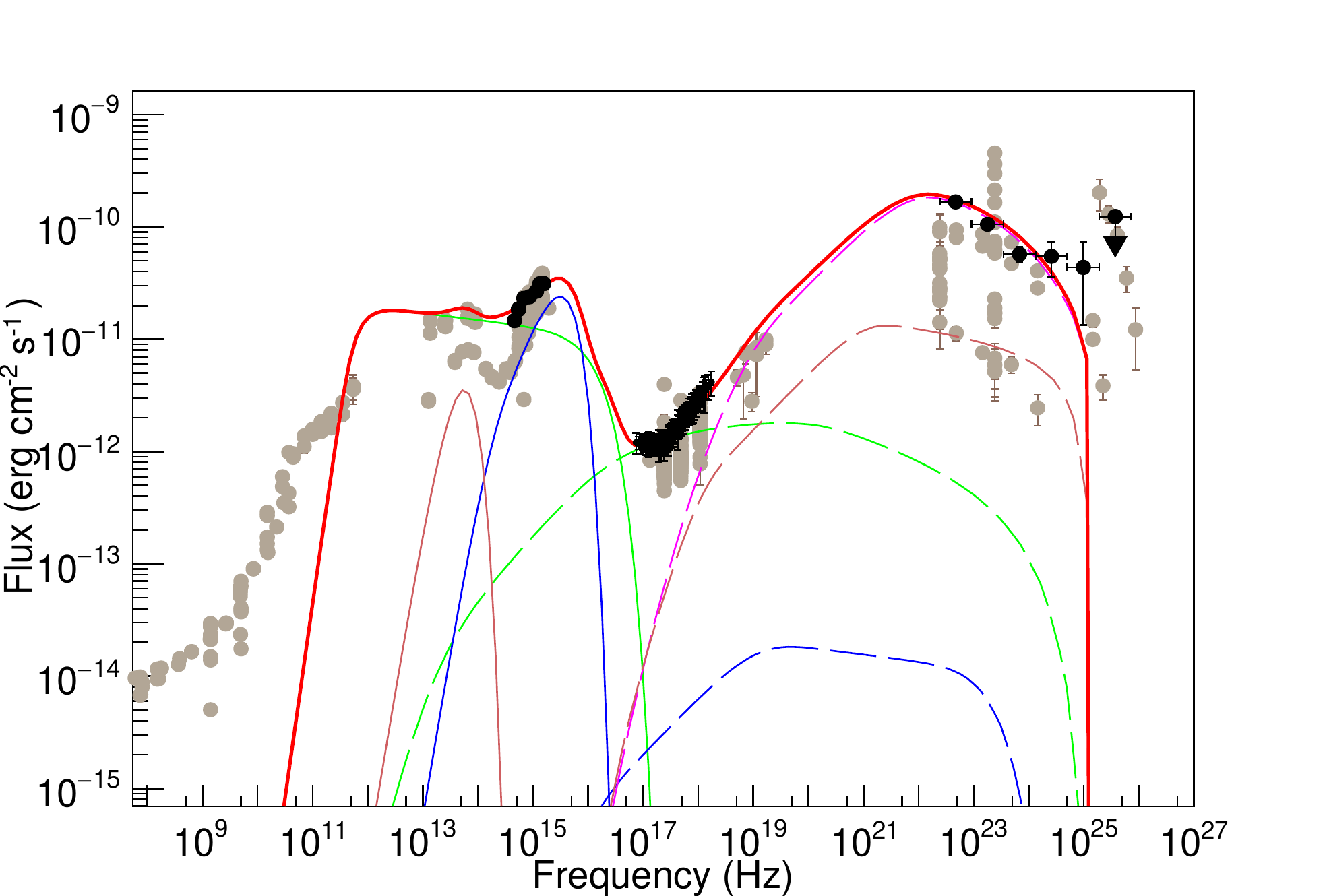} & 
\includegraphics[width=0.5\textwidth]{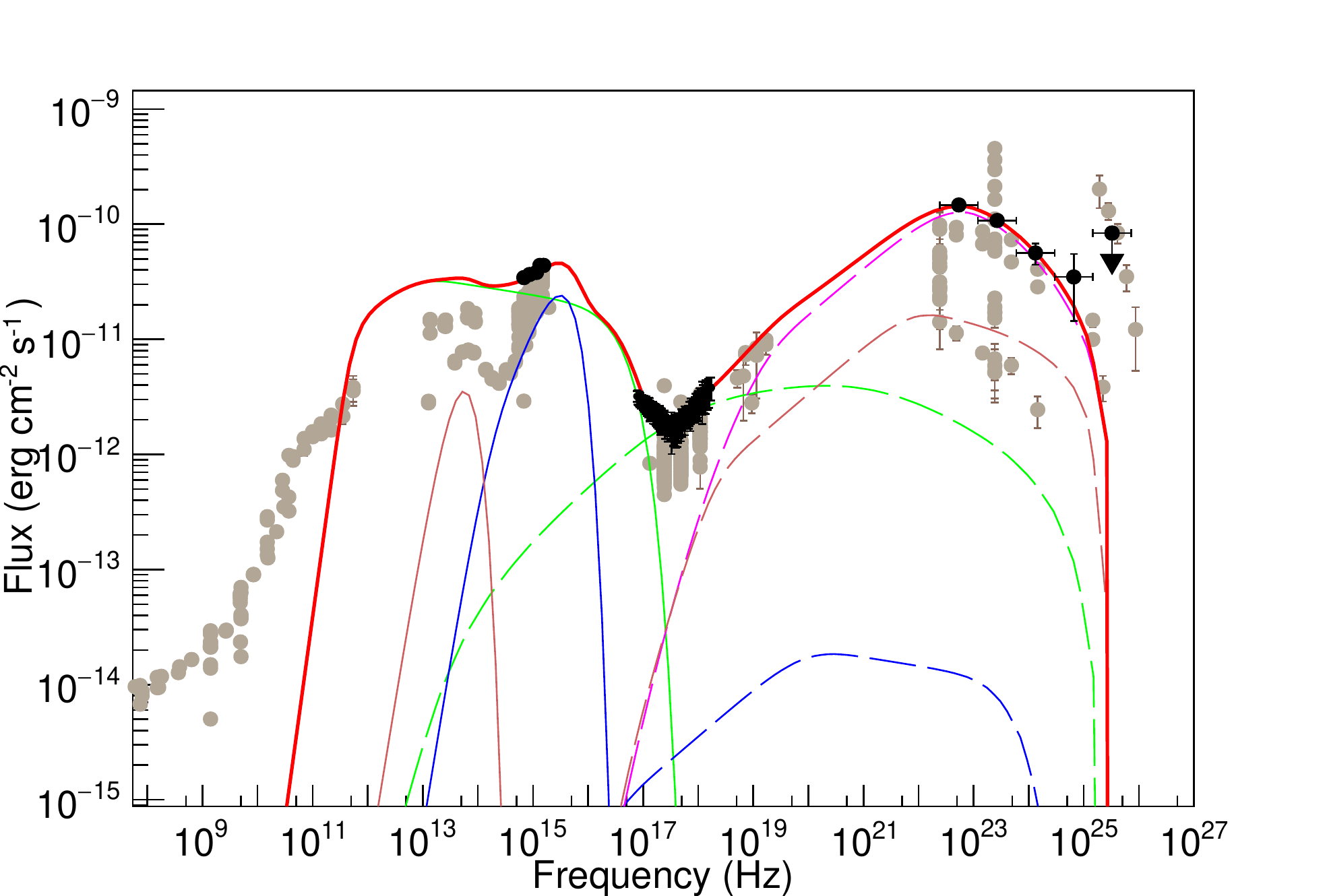} \\
\textbf{(A)}  & \textbf{(B)}  \\[6pt]
\end{tabular}
\begin{tabular}{cc}
\includegraphics[width=0.5\textwidth]{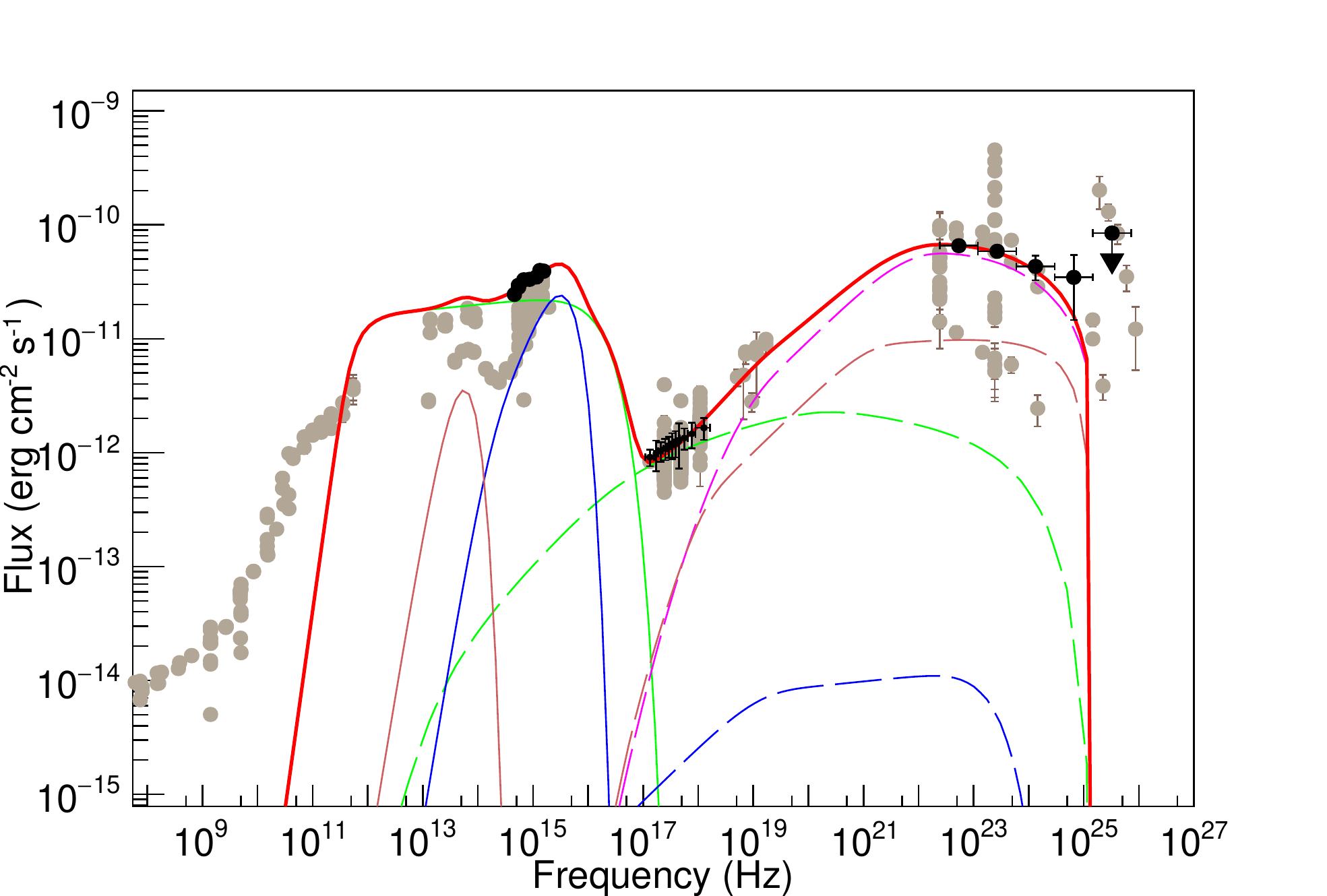} &
\includegraphics[width=0.5\textwidth]{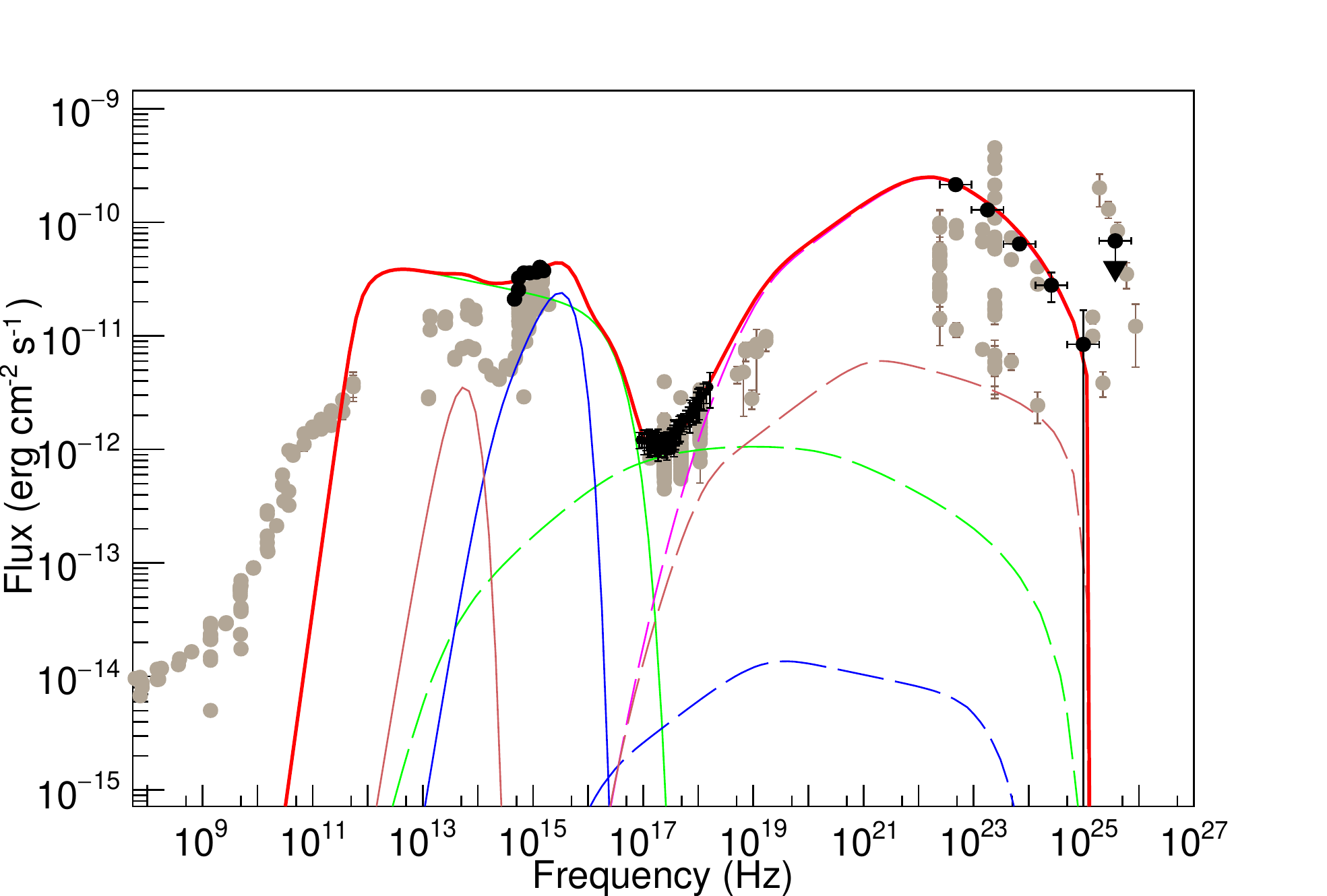} \\
\textbf{(C)}  & \textbf{(D)}  \\[6pt]
\end{tabular}
\caption{The fitted broadband SEDs of both Flare-A and B. \textbf{Panel A}: Flare-A Block-1 (MJD 56680-56700) SED. \textbf{Panel B}: Flare-A Block-2 (MJD 56700-56725) SED. \textbf{Panel C}: Flare-A Block-3 (MJD 56725-56748) SED. \textbf{Panel D}: flare-B SED. In each panel, the solid red curve represents the final fitted model, and the other curves show its different components like synchrotron emission (green solid curve), SSC (dashed green curve), dusty torus (DT) emission (brown solid curve), EC-DT emission (brown dashed curve), disk thermal emission (blue solid curve), EC-disk emission (blue dashed curve) and EC-BLR emission (pink dashed line). The black points represent multi-waveband data and upper limits with the corresponding errors. The grey points represent the archival data from earlier studies.}
\label{fig:SED_fit}
\end{figure*}

Panel A, B, and C of \autoref{fig:SED_fit} represent the SED of three blocks of Flare-A fitted with a single-zone leptonic model. The details of the fitted parameters are given in \autoref{tab:fitted_par}. In Block-1, the X-ray part is modelled with synchrotron, SSC and EC-BLR components, whereas the $\gamma-$ray data is dominated by the EC-BLR component. The comparatively low flux in the optical-UV band justifies the low value of the magnetic field. 
To fit the X-ray data with SSC and falling part of synchrotron hump, $N$, $\gamma_{min}$, and $\gamma_{max}$ are varied within accessible parameter space. The comparatively high value of $N$ compensates for the effect of the lower magnetic field in producing the SSC component.
In Block-2, the typical X-ray spectral shape suggests that the low energy part (i.e. below $E_b$) falls into the falling portion of the first hump. Thus the low energy X-ray, up to $E_b$, is the result of synchrotron emission. The production of keV photons in synchrotron emission requires either a strong magnetic field or injection spectrum with a good fraction of high energy particles. Enhancement of magnetic field to its maximum value, allowed by optical-UV points, is not alone sufficient. Thus we start scanning the available parameter space of $E_b$, and $\gamma_{max}$ until the low energy X-ray part fits well.
To model the X-ray spectra above $E_b$ with the SSC component, $\gamma_{min}$ is slightly increased from the Block-1 value. The parameter $p_1$ is adjusted to maintain the slope of the rising part of the second hump. From the final values of jet parameters, it is observed that the overall shift of the input particle spectrum towards the high energy regime can explain the SED of Block-2. In Block-3, flattening of Fermi-LAT data provides a low value of $p_2$, which along with $\gamma_{min}$ increases the SSC components to fit the X-ray data. The strength of the EC-BLR component in producing the second hump gradually decreases from Block-1 to Block-3, as explained in the discussion.
\begin{table*}
	\centering
	\caption{SED model parameters for one zone leptonic scenario. External parameters remain the same for both flare states.}
	\label{tab:fitted_par}
	\begin{tabular}{ccccc} 
		\hline
		Jet parameter & \multicolumn{3}{c}{Flare-A} & flare-B \\
		\cline{2-4}
		& Block-1 & Block-2 & Block-3 & \\
		\hline
		\hline
		$R$($10^{16}$ cm) & 1.724 & 1.724 & 1.724 & 2.055 \\
		$d(10^{17}$ cm) & 8.0 & 8.5 & 9.0 & 7.0 \\
		$B(G)$ & 1.22 & 1.44 & 1.4 & 2.65 \\
		$N$($\text{cm}^{-3}$) & 8900 & 6000 & 6000 & 5000 \\
		$\gamma_{min}$ & 3 & 4 & 3 & 3 \\
		$\gamma_{max}$ & 13000 & 20000 & 14000 & 12000 \\
		$\gamma_b$ & 110 & 300 & 120 & 115 \\
		$p_1$ & 1.99 & 2.18 & 2.06 & 2.26 \\
		$p_2$ & 3.12 & 3.15 & 2.9 & 3.2 \\
		\hline
		External parameter & & & & \\
		\hline
		Accretion rate $(\text{M}_{\odot}/\text{yr})$ & & 7.718 & & \\
		$R_{BLR}^{in}$ ($10^{17}$ cm) & & 5.816 & & \\
		$R_{BLR}^{out}$ ($10^{17}$ cm) & & 6.016 & &  \\
		$\tau_{\text{BLR}}$ & & 0.1 & & \\
		$R_{DT}$ ($10^{18}$ cm) & & 14.79 & & \\
		$T_{DT}$ (K) & & 1000 & & \\
		$\tau_{DT}$ & & 0.1 & & \\
		\hline
	\end{tabular}
\end{table*}
\subsubsection{Flare-B (MJD 56950--57000)}
Due to the lack of flux variation in both optical-UV and X-ray bands, we have modelled Flare-B as a single SED instead of splitting into Blocks. In the optical band, a total of 8 observations are available from SPOL-CCD instruments during Flare-B. The average flux for both V and R band filters after reddening correction are given in \autoref{tab:opt_details}. The coverage of \textit{Swift} data is comparatively less during this flare. The \textit{Swift}-UVOT and XRT instruments provide only $3$ observations in UV and X-ray bands.
\autoref{tab:uvot_details} provides the average optical and UV fluxes for all six filters after the corresponding reddening corrections.
The results from X-ray data analysis are given in \autoref{tab:xrt_details}. As in previous cases, fluxes are corrected for the galactic soft X-ray absorption.
The results from analysis of $\gamma-$ray data are given in \autoref{tab:lat_details}. Based on 4FGL catalog, $\gamma-$ray data are fitted using log-parabola model with fixed break-energy ($E_b$). 

The right panel of \autoref{fig:Obs_SED} shows broadband SED of Flare-B. The optical-UV band shows a comparatively flat spectrum. The X-ray data shows the same characteristic feature as Block-1 of Flare-A. It suggests that the X-ray emission is the combination of synchrotron and SSC components. The high energy part of the X-ray spectrum and $\gamma-$ray emission forms the second hump. The modelling approach adopted was similar to Block-1 of Flare-A. The blob position ($d$), magnetic field, and the other jet parameters were varied to fit the SED by eye estimation. Panel D of \autoref{fig:SED_fit} represents the final SED of flare-B fitted with a single zone leptonic model. 
The SPOL data points are not used in this SED fitting. The details of fitted parameters are given in \autoref{tab:fitted_par}. The presence of the emission region close to BLR makes EC-BLR the only dominant component in the second hump.

\subsection{Jet Power Estimation}
The power associated with a relativistic jet ($P_{Jet}$) is the sum of the power of relativistic leptons, cold protons, and the entangled magnetic field. From the physical parameters of the emission region obtained from the SED modelling, we can estimate the power associated with different components and it is given by,
\begin{equation}
    P_i = 2\pi R^2\Gamma^2\beta c U_i
\end{equation}
where $U_i$ represents the i$^{\text{th}}$ component energy density in the comoving frame and factor of 2 includes the effect of both forward and backward jet. Thus power associated with relativistic leptons is given as,
\begin{equation}
    P_e = 2\pi R^2 \Gamma^2 \beta c m_e c^2 \int_{\gamma_{min}}^{\gamma_{max}}n(\gamma)\gamma d\gamma= 2\pi R^2 \Gamma^2 \beta m_e c^3 n_e\langle \gamma \rangle 
\end{equation}
Assuming the presence of cold protons, with number density $n_p$ in the jet, associated power is given by,
\begin{equation}
    P_p = 2\pi R^2\Gamma^2\beta m_p c^3 n_p
\end{equation}
In the case of the magnetic field, this is given by,
\begin{equation}
    P_B = 2\pi R^2 \Gamma^2 \beta c U_B
\end{equation}
A good fraction of jet power is spent to produce the observed non-thermal emission which is given by an almost model-independent formula-
\begin{equation}
    P_r = L \frac{\Gamma^2}{4\delta^4}
\end{equation}
where L is the total observed non-thermal luminosity \citep{Ghisellini2010MNRAS}. The estimated power of each component is given in \autoref{tab:pow_estimation}.
\begin{table}
	\centering
	\caption{Estimated values of power associated with different components of jet}
	\label{tab:pow_estimation}
	\scalebox{0.98}{
	\begin{tabular}{lcccc} 
		\hline
		Jet parameter & \multicolumn{3}{c}{Flare-A} & flare-B \\
		\cline{2-4}
		& Block-1 & Block-2 & Block-3 & \\
		\hline
		\hline
		$U_e$(erg/cc) & 1.01e-2 & 7.9e-2 & 6.6e-2 & 4.2e-2 \\
		$U_B$(erg/cc) & 5.9e-2 & 8.2e-2 & 7.8e-2 & 2.8e-1 \\
		$P_e$(erg/s) & 1.5e+45 & 1.2e+45 & 9.8e+44 & 8.7e+44 \\
		$P_p$(erg/s) & 2.0e+46 & 1.3e+46 & 1.3e+46 & 1.6e+46 \\
		$P_B$(erg/s) & 8.7e+44 & 1.2e+45 & 1.2e+45 & 5.9e+45 \\
		$P_{rad}$(erg/s) & 5.5e+44 & 4.9e+44 & 2.8e+44 & 7.4e+44 \\
		$P_{Jet}$(erg/s) & 2.2e+46 & 1.5e+46 & 1.5e+46 & 2.3e+46 \\
		\hline
	\end{tabular}
	}
\end{table}
\section{Discussion}
\label{sec:dis}
In this work, we have carried out a detailed study of two flares of PKS B1222+216 observed in the first and fourth quarter of $2014$ which includes temporal study and broadband SED modelling with the leptonic scenario. As a part of the temporal study, we have searched for possible flux-index correlation in the $\gamma$-ray band for both flares (\autoref{fig:flux-index corr}). We have estimated the Pearson correlation coefficient for Flare-A. For complete data set, a weak correlation of 0.24 is obtained whereas a significant improvement in correlation coefficient can be achieved by excluding data points above flux $10^{-6}\text{ ph}\text{ cm}^{-2}\text{s}^{-1}$ ($20\%$ of complete data set). The corresponding correlation coefficient is 0.56. It implies a possibility of a strong flux-index correlation in the low flux state of Flare-A that disappears during the six-day long high flux period. These two ranges are also fitted with a straight line which shows \lq brighter when softer{\rq} scenario.
It is contrary to the trend mentioned in \citet{Bhattacharya2021MNRAS}. Flare-B does not show any such correlation in the flux-index plot. \autoref{fig:flux-index hys} shows spectral evolution of source over a short period around high flux state during both the flares. In the case of Flare-A, the source exhibits an anti-clockwise pattern which implies a steepening in the observed photon spectrum during the peak flux state. It is the result of the energy-dependent cooling process and rapid change in the injection spectrum. Due to a higher cooling rate, the high energy counterpart of the emitting lepton population decays faster than the low energy part that makes it steeper. Thus the changes in particle injection first affect the high energy counterpart of the emitting population, and finally, this information propagates to the lower energy part \citep{Tashiro1995PASJ, Kirk1998A&A}. The emitted photon spectrum related to this lepton population reflects the same trend. The same pattern had previously been reported in the 2010 flare of PKS B1222+216 \citep{Kushwaha2014ApJ, Tanaka2011ApJ}. It also implies that the flare evolution is primarily controlled by the cooling mechanism and injection of low energy particles. In the case of Flare-B, no such specific trend is obtained \citep{Kushwaha2014ApJ}. It suggests that the triggering of Flare-A may be primarily started with an injection of a low energy particle ($< 0.3$ GeV) population.

We have systematically studied Fermi-LAT LC by dividing it into three consecutive bands (E1: $0.1-0.3$ GeV, E2: $0.3-1$ GeV, and E3: $1-300$ GeV) and analyzed it with the shortest possible time bin of 12 hours (\autoref{fig:flare_fitting}). In contrast to the combined ($0.1-300$ GeV) LC in the upper panel, the individual bands show some complex temporal behaviour. In the E1 band, Flare-A shows two subflares which are separated by $\sim$3 days. Starting with a slowly rising trend, subflare-1 reaches its maximum at MJD $(56699\pm0.5)$, followed by a rapid descent. On the other hand, subflare-2 increases rapidly to its maximum (MJD $56702\pm0.5$) and follows a slow decaying path. The emissions in the other two bands (E2 and E3) are solely governed by subflare-2. The highest flux observed during the course of Flare-A is $\text{F}_{0.1-300\text{ GeV}}=(3.2\pm0.42)\times10^{-6}\text{ ph}\text{ cm}^{-2}\text{s}^{-1}$ at MJD $(56702.25\pm0.5)$. In combined LC, the observed plateau portion, just before the peak flux, is the contribution of subflare-1 which justifies its absence in E2 and E3 bands. The absence of the plateau in HR1 and HR2 in the left panel of \autoref{fig:Hardness_ratio} supports the above statement. The occurrence of the plateau phase before the flare was previously observed in blazar 3C 454.3 \citep{Abdo2011ApJ} and 2010 flare of PKS B1222+216. 

Symmetric temporal evolution, with an equal rise and decay times, may occur when a perturbation in the jet flow or a blob of denser plasma passes through a standing shock present in the jet \citep{Blandford1979ApJ}. The subflare-1 has a longer rise time than decay time, which can happen due to a slow injection rate in the emission region. On the other hand, the long decay time of subflare-2 in the E1 band could result from either a longer cooling time of leptons or a weakening of the acceleration mechanism. Since LC fitting can only provide information about physical processes which are slower than the duration of the event \citep{Chiaberge1999MNRAS, Chatterjee2012ApJ}, we can only conclude that starting with a slow injection, subflare-1 passed through a faster cooling phase, whereas the opposite happened in subflare-2.


Flare-B shows two distinct subflares in all the three energy bands which are also separated by a time span of $\sim$3 days. In E1 and E2 band, both these subflares are characterized by a rapid rise time ($\sim$0.3 days) followed by a comparatively long decay time ($\sim$0.6 days). In E3 band, subflare-2 follows an opposite trend with long rise time and shortest decay time ($\sim0.2$ days). The highest flux obtained during the course of flare-B is $\text{F}_{0.1-300\text{ GeV}}=(6.93\pm0.64)\times10^{-6}\text{ ph}\text{ cm}^{-2}\text{s}^{-1}$ at MJD $(56973.25\pm0.5)$. \autoref{tab:flare_fit_par} shows that both the subflares are characterized by very short rise time (average of 0.34 days) in all energy bands. This could suggest the presence of a short-lived acceleration like first order Fermi acceleration mechanism with acceleration timescale $t_{acc}\sim(r_g/c)(c/u_s)^2$, where $u_s$ is the speed of shock and $r_g$ is the electron gyroradius \citep{Kirk2001JPhG,Protheroe2004PASA,Rieger2007Ap&SS}.

One interesting feature of flare-B is the gradual change in relative flux strength of subflares in each band (\autoref{fig:flare_fitting}). Moving from E1 to E3 band, the ratio of peak flux of subflare-1 to subflare-2 changes from $1.79$ to $0.63$. This implies that subflare-1 is primarily dominated by low energy particles whereas subflare-2 is dominated by high energy counterpart. This is also justified from the right panel of \autoref{fig:Hardness_ratio}. The HR2 plot shows the local minimum and maximum corresponding to subflare-1 and subflare-2, respectively. This also indicates that flare-B is primarily triggered with an injection of low energy particles and later high energy counterpart comes into play. The closeness of the rise and decay times for both subflares in the E1 and E2 band (\autoref{tab:flare_fit_par}) and the almost constant nature of HR1 (right panel of \autoref{fig:Hardness_ratio}) indicates a strong correlation between E1 and E2 band whereas correlation with E3 band is comparatively less.

In the case of Block-1 (Panel A, \autoref{fig:SED_fit}), the thermal component significantly surpasses non-thermal emission in the optical-UV range. This can be characterized as a signature of the pre-flare stage. The shifting of break-energy ($\gamma_{\text{b}} = 300$) in Block-2, lowering of $p_2$ ($p_2$ = 2.9) in Block-3 and higher magnetic field ($B$ = 3.2 G) in flare-B enhances synchrotron emission and makes it comparable to thermal counterpart. Due to the fixed value of SMBH mass ($\text{M}_{\text{BH}}$) and disc luminosity ($\text{L}_\text{D}$), thermal component remains same for all SEDs.

According to the standard FSRQ model, the high energy part of the SED second hump is the contribution of the EC emission mechanism where reprocessed thermal photons from both BLR and dusty torus (DT) region take part as seed photons. The density of seed photons inside the emission region primarily determines the fractional contribution of both EC components in constructing the second hump. Due to this, when the blob is closer to BLR than the dusty torus region, EC-BLR component dominates over EC-DT component. This feature is clearly seen in our SEDs (\autoref{fig:SED_fit}). In Flare-A, the emission region gradually moves away from the BLR as we go from Block-1 to Block-3. It decreases the density of reprocessed photons from BLR inside the emission region
and the corresponding reduction in the fractional contribution of the EC-BLR component is visible in panel A, B, and C of \autoref{fig:SED_fit}. Despite the increasing distance from BLR, the blob position is still far away from the dusty torus region ($d/R_{DT}\sim0.061$) which justifies the nearly constant nature of the EC-DT component throughout Flare-A. In the case of Flare-B, the emission region is situated just outside the BLR and far away from DT. Despite being Doppler de-boosted in blob rest-frame, a comparatively long distance between blob and a dusty torus is the main reason for the lower value of EC-DT compared to the EC-BLR component.

\begin{figure}
\centering
\includegraphics[width=0.5\textwidth]{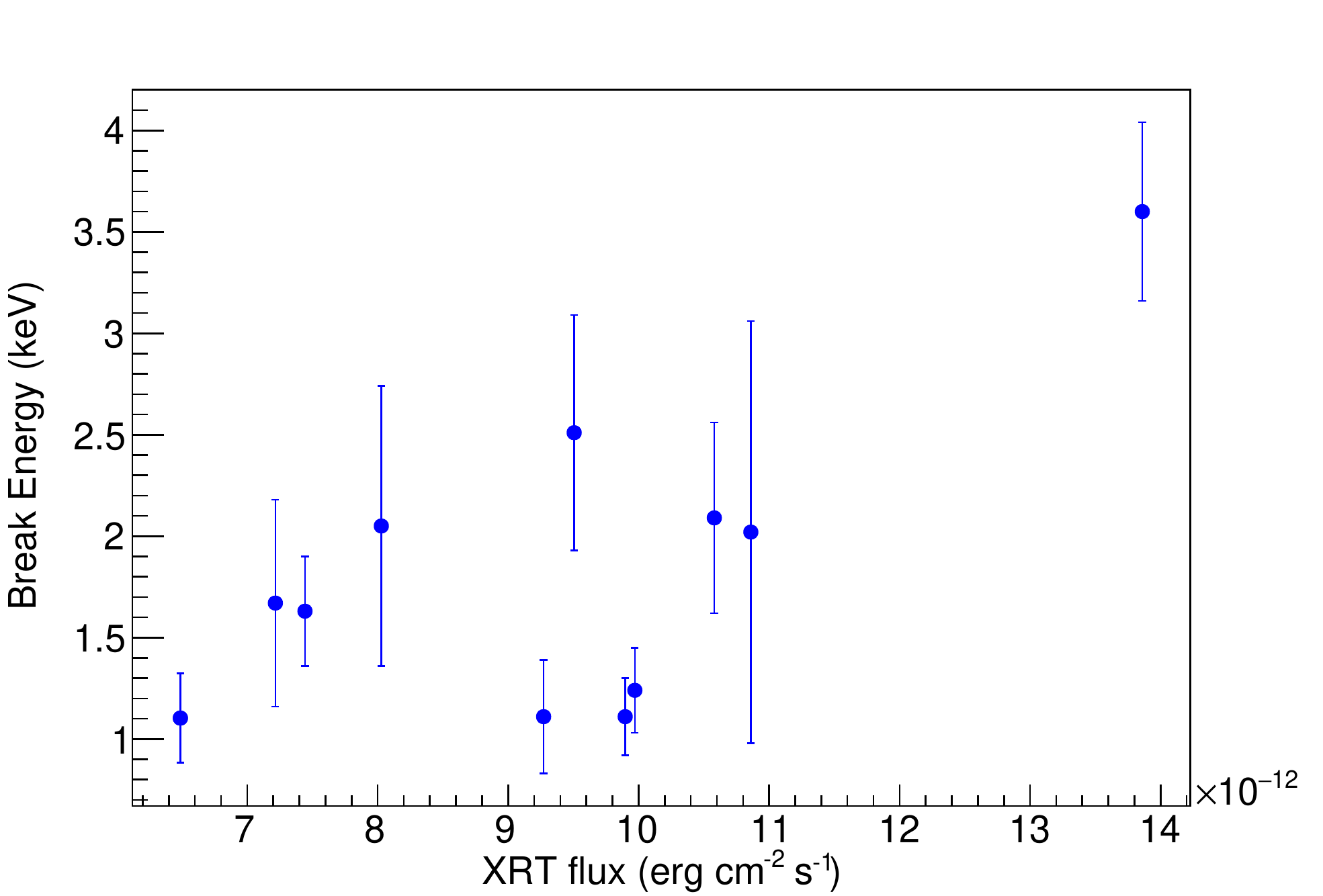}
\caption{Variation of break-energy ($E_b$) with XRT integrated flux over 0.3-8.0 keV.}
\label{fig:XRT-flux-vs-Eb}
\end{figure}

\begin{figure}
\centering
\includegraphics[width=0.5\textwidth]{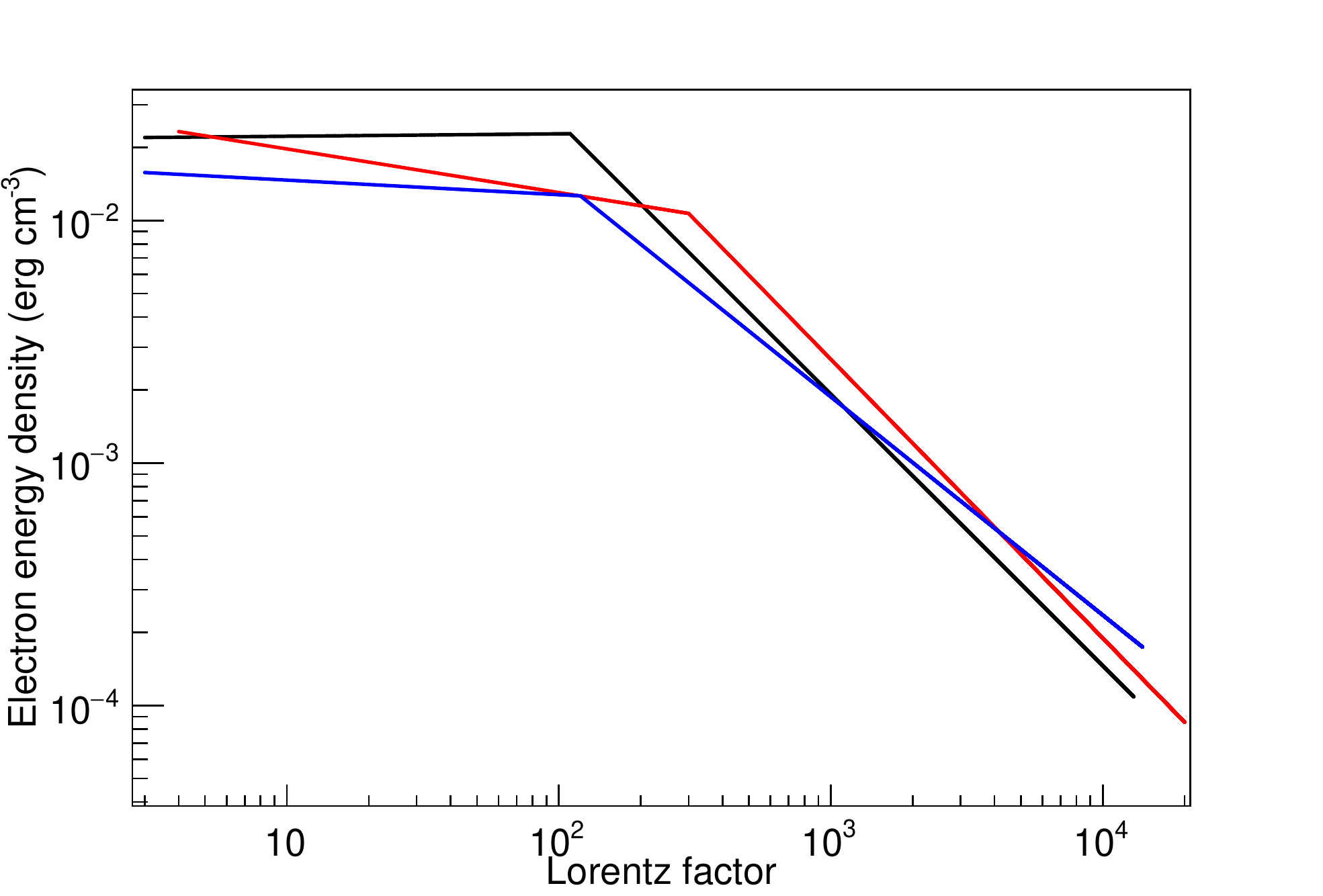}
\caption{The evolution of lepton distribution (energy density as a function of $\gamma$) during  Flare-A is shown. The black, red, and blue curves represent particle distributions corresponding to Block-1, Block-2, and Block-3 SED respectively.}
\label{fig:Evolution_lepton_distribution}
\end{figure}

In Flare-A and Flare-B, one of the most interesting features is the evolution of the X-ray spectrum during the flaring activity which is not reported previously. To explore this feature, we have analyzed all individual X-ray observations with XSPEC and tried to find out the best spectral form (power-law or broken power-law) to fit the data. In the case of Flare-A, out of a total of $18$ observations, only $10$ observations reveal their broken power-law spectral nature. Except for one (Obs ID: 00036382033) from Block-1, all these observations fall in the period of Block-2 (MJD 56700-56725). In the case of Flare-B, one observation (Obs ID: 00036382060) out of 3 observations shows the same spectral nature. The break-energy features are not present in the spectra of other observations. Some observations are well fitted with power-law and for others, the exposure time is not sufficient to reveal this nature. The details of best-fitted parameters are given in \autoref{tab:xrt_indivi_analy}. The break-energy represents the location in the broadband SED above which SSC emission dominates over synchrotron. We have studied the variation of break-energy with integral flux over the energy range $0.3$--$8.0$ keV (\autoref{fig:XRT-flux-vs-Eb}). The correlation between flux and $E_b$ is studied in terms of the Pearson correlation coefficient which is found to be $0.7$. This suggests a strong correlation between the increasing trend of break-energy with the increase of corresponding integral flux. This can suggest a possible explanation for the evolution of X-ray spectra. 
In Block-3, a comparatively low flux state (or low count rate) constrain the break-energy somewhere below $0.3$ keV which provides a simple power-law spectrum in the energy range of $0.3-8.0$ keV.  
This is also consistent with usual FSRQ SEDs where break-energy i.e. the transition point between the first and the second hump, falls between UV and soft X-ray regime. The lack of observational data over this range stops us from concluding anything further. Comparatively high flux (or count rate) in Block-1 pushes the transition point above $0.3$ keV and consequently forms a broken power-law spectral shape with low break-energy (0.98 keV). With the rapid enhancement of flux (or count rate) in Block-2, the break-energy increases well above 0.3 keV (1.46 keV) and produces a characteristic broken power-law spectrum in the soft X-ray band. In other words, the lower hump tail (decaying part) of SED is shifted towards the high-frequency regime which affects the X-ray spectrum below $E_b$ and abruptly changes the photon spectral index from $2.02$ to $2.459$. On the other hand, this change does not affect the Block-2 X-ray spectra above $E_b$ which is evident from the closeness of X-ray spectral index $P_2$ in Block-1 and Block-2 (\autoref{tab:xrt_details}).
The rapid change in the photon spectrum generally suggests either passage of strong shock through the emission region or rapid change in injection spectrum that affects the complete particle spectrum inside the blob. It is observed from \autoref{tab:fitted_par} that the Block-2 particle distribution primarily differs from others in terms of higher values of $\gamma_{\text{min}}$, $\gamma_{max}$ and $\gamma_b$. The shift of break Lorentz factor ($\gamma_b$) in Block-2 to a higher value implies the enhancement of the low energy lepton density, i.e. an increment in the fraction of particle density below $\gamma_b$ inside the emission region. This could be the result of either a rapid injection of low energy particles or a strengthening of the acceleration mechanism which only affects the low energy population. This change in emitting particle population boosts up the optical-UV emission and extends the first hump up to a soft X-ray regime. The abrupt change in the low energy population does not affect the high energy $\gamma$-ray emission as observed from nearly the same spectral nature in Block-1 and Block-2. The particle spectrum, responsible for emission in each block, is schematically shown in \autoref{fig:Evolution_lepton_distribution}. 

The observation of PKS B1222+216 in the very high energy $\gamma$-ray band with the shortest variability timescale of $\sim10$ minutes challenges standard emission models by constraining the size of the emission region ($R$ $\sim5\times10^{14}$ cm). In \citet{Tavecchio2011A&A}, SED modelling of 2010 flares with both one and two emission regions and their relative positions in the jet to fit the observed data had been studied. The combination of a compact blob outside the BLR and a larger blob inside the BLR can explain the observed SED. \citet{Kushwaha2014MNRAS} proposed an emission model where broad-band emission originates from a compact region, arising plausibly from the compression of jet matter at the recollimation zone, and finally flare is interpreted as an outcome of jet deceleration in the shock. In \citet{Kushwaha2014ApJ}, detailed spectral behaviour was studied by dividing the Fermi energy range into four sub-energy bands. The study of flux-index correlation showed a similar \lq softer when brighter \rq scenario. The detailed flare fitting in various sub-bands provides information about the underlying acceleration and cooling mechanism. In \citet{Bhattacharya2021MNRAS}, the flaring activity in 2014 is considered as the medium flux (MS) state and SED modelling is only performed for Flare-B with a duration of $\sim$7 days. In spite of the leptonic model scenario, the modelling approach is different from ours. In \citet{Abhradeep2021MNRAS}, a similar one-zone leptonic model (using JETSET code) has been used to describe the quiescent state broadband SED of PKS B1222+216. In the present work, we have found a comparatively harder Fermi spectral index ($\alpha$) than that of the quiescent state as expected.
\begin{table*}
	\centering
	\caption{Details of \textit{Swift}-XRT individual observation analysis. Observations are fitted with the broken power law model. The best fitted value of its parameters like first/second spectral index ($P_1$/$P_2$), break-energy ($E_b$) and normalization constant ($K$) and photon flux are given below. }
	\label{tab:xrt_indivi_analy}
	\begin{tabular}{ccccccc} 
		\hline
		 Obs ID & $P_1$ & $P_2$ & $E_b$ & $K$ & $\chi_{r}^2$ & $\text{F}_{0.3-8.0\text{ keV}}$ \\
		  & & & (keV) & ($10^{-4}\text{ ph}\text{ cm}^{-2}\text{s}^{-1}\text{keV}^{-1}$) & & ($10^{-12}\text{ erg}\text{  cm}^{-2}\text{s}^{-1}$) \\
		\hline
		\hline
		00036382033 & $2.74\pm0.31$ & $1.64\pm0.22$ & $1.11 \pm 0.28$ & $12.26 \pm 2.06$ & $0.55$ & $9.27$\\
		00036382040 & $2.82 \pm 0.16$ & $1.03 \pm 0.32$ & $1.63 \pm 0.27$ & $9.29 \pm 0.82$ & $0.43$ & $7.44$\\
		00036382045 & $2.35 \pm 0.15$ & $0.1 \pm 1.13$ & $2.05 \pm 0.69$ & $11.41 \pm 0.08$ & $0.37$ & $8.03$\\
		00036382046 & $2.60 \pm 0.29$ & $0.1 \pm 2.45$ & $2.02 \pm 1.04$ & $1.0 \pm 0.12$ & $0.08$ & $10.86$\\
		00036382048 & $2.37 \pm 0.16$ & $1.48 \pm 0.29$ & $1.67 \pm 0.51$ & $11.31 \pm 0.83$ & $0.46$ & $7.22$\\
		00036382049 & $2.4 \pm 0.13$ & $0.58 \pm 0.71$ & $2.51 \pm 0.58$ & $13.67 \pm 0.85$ & $0.95$ & $9.51$\\
		00036382050 & $2.92\pm 0.26$ & $1.27\pm 0.17$ & $1.11 \pm 0.19$ & $9.78 \pm 1.51$ & $0.88$ & $9.9$\\
		00036382051 & $2.07\pm 0.14$ & $0.83\pm 0.39$ & $2.09 \pm 0.47$ & $1.31 \pm 0.08$ & $0.71$ & $10.58$\\
		00036382052 & $2.12\pm 0.1$ & $-1.53\pm 1.33$ & $3.6 \pm 0.44$ & $13.72 \pm 0.67$ & $0.77$ & $13.86$\\
		00036382053 & $2.67\pm 0.19$ & $1.67\pm 0.16$ & $1.24 \pm 0.21$ & $14.55 \pm 1.38$ & $1.37$ & $9.97$\\
		00036382060 & $2.45\pm 0.3$ & $1.26\pm 0.14$ & $1.1 \pm 0.22$ & $6.76 \pm 1.06$ & $0.26$ & $6.488$\\
		\hline
	\end{tabular}
	\end{table*}
\section{Conclusions}
\label{sec:con}
This study aimed to carry out an in-depth temporal and spectral analysis of PKS B1222+216 during the flaring activities in the first and fourth quarters of 2014. To perform this study, we have analyzed the optical/UV, X-ray, and $\gamma-$ray data and collected optical data from the SPOL-CCD archive for these periods. We present a detailed study of $\gamma-$ray data which includes flux-index correlation study, modelling of flares, and hardness ratio test for three energy bands. Due to rapid flux variation, Flare-A divides into three consecutive Blocks. We have used a time-dependent leptonic jet model with an internal shock scenario (JETSET) to reproduce the above-mentioned spectral states of PKS B1222+216. Though the presented model parameters are not a unique set to reproduce the observed data, they provide a well-constrained parameter space for the values of key physical parameters using quasi-simultaneous multi-wavelength data. Our conclusions are as follows:

(i) A relatively high flux-index correlation in the low flux state and corresponding hysteresis loop in the high flux state of Flare-A suggest a flaring activity, controlled by cooling mechanisms and injection of low energy particles, within \lq softer when brighter{\rq} scenario of low flux state.

(ii) The presence of plateau just before the peak flux state suggests that Flare-A might be triggered with a rapid injection of low energy particle $(<0.3\text{ GeV})$ followed by high energy particle population. In Flare-B, the decaying trend in the amplitude of subflare-1 and the corresponding rising trend in subflare-2 also suggest the same scenario as in Flare-A. 

(iii) The consistent short rise time of flare-B generally indicates the presence of a short-lived acceleration like the first-order Fermi acceleration mechanism.

(iv) In the soft X-ray regime, a correlation between the break-energy and the integral flux is observed. The enhancement in the observed integral flux (0.3 - 8.0 keV) is the consequence of the increase in break-energy.

(v) One-zone leptonic model with a broken power-law lepton distribution is sufficient to explain the observed features of both the flares of 2014. In both Flare-A and Flare-B, the shifting of the synchrotron emission profile towards the higher energies regime is observed and interpreted as the direct consequence of a shift in the break-energy of the emitting lepton population. The analysis of the Fermi LC is also consistent with this interpretation.

\section*{Acknowledgements}
In this work, we have used data from the Steward Observatory that is part of a spectropolarimetric monitoring project. This program is supported by Fermi Guest Investigator grants NNX08AW56G, NNX09AU10G, NNX12AO93G, and NNX15AU81G. We have used data from both the XRT and UVOT instruments of Neil Gehrels Swift Observatory. This research has made use of the XRT Data Analysis Software (XRTDAS) developed under the responsibility of the Space Science Data Center (SSDC) maintained by the Italian Space Agency (ASI). We have used Fermi-LAT data, obtained from the
Fermi Science Support Center, provided by NASA’s Goddard Space Flight Center (GSFC). The data and analysis software were obtained from NASA’s High Energy Astrophysics Science Archive Research Center (HEASARC), a service of GSFC. We used a community-developed Python package named Enrico to make Fermi-LAT data analysis easier and more convenient \citep{Sanchez2013arxiv}. Finally, We acknowledge the financial support of the Department of Atomic Energy (DAE), Government of India, under Project Identification No. RTI 4002.
\section*{Data Availability}
(i) The SPOL-CCD data used in this article
are available in the Steward Observatory Data Archive at \url{http://james.as.arizona.edu/~psmith/Fermi/}.

\hspace{-0.32cm}(ii) The Swift-XRT data used in this article
are available in the NASA's HEASARC Archive at \url{https://heasarc.gsfc.nasa.gov/docs/archive.html}.

\hspace{-0.32cm}(iii) The Fermi-LAT data used in this article
are available in the LAT data server at \url{https://fermi.gsfc.nasa.gov/ssc/data/access/}.

\hspace{-0.32cm}(iv) The Fermi-LAT data analysis software is available at \url{https://fermi.gsfc.nasa.gov/ssc/data/analysis/software/}.

\hspace{-0.32cm}(v) The software used for Broadband SED modelling is available at \url{https://github.com/andreatramacere/jetset}.

\hspace{-0.32cm}(vi) The archival data underlying this article
are available in the SSDC (ASI) archive at \url{https://www.ssdc.asi.it/}.

\hspace{-0.32cm}(vii) The data derived in this article will be shared on reasonable request to the corresponding author.



\bibliographystyle{mnras}
\bibliography{reference} 








\bsp	
\label{lastpage}
\end{document}